\documentclass[twocolumn]{aastex631}

\newcommand{\mace}{\textsc{mace}}

\newcommand{\Phantom}{\textsc{Phantom}}
\newcommand{\csone}{\textsc{1d-cse}}
\newcommand{\cszero}{\textsc{0d-cse}}

\newcommand{\Msol}{\rm{M_{\odot}}}

\newcommand{\kms}{\rm{km\,s^{-1}}}
\newcommand{\Msolyr}{\Msol\,yr^{-1}}

\newcommand{\cm}{{\rm cm}}

\newcommand{\Tstar}{T_\star}

\newcommand{\Mdot}{\dot{M}}

\newcommand{\Rstar}{R_\star}

\newcommand{\vexp}{v_{\rm exp}}
\newcommand{\eps}{\varepsilon}
\newcommand{\n}{\boldsymbol{n}}
\newcommand{\p}{\boldsymbol{p}}
\newcommand{\z}{\boldsymbol{z}}
\newcommand{\x}{\boldsymbol{x}}
\newcommand{\A}{\boldsymbol{\mathcal{A}}}
\newcommand{\B}{\boldsymbol{\mathcal{B}}}
\newcommand{\C}{\boldsymbol{\mathcal{C}}}
\newcommand{\D}{\mathcal{D}}
\newcommand{\E}{\mathcal{E}}
\newcommand{\AV}{A_{\rm V}}

\usepackage{amsmath}
\usepackage{amssymb}
\usepackage{bbold}
\usepackage{latexsym}
\usepackage{graphicx}
\usepackage{subcaption, caption}
\usepackage{natbib}
\usepackage[section]{placeins}

\begin{document}

\title{{MACE}: A Machine learning Approach to Chemistry Emulation}

\correspondingauthor{Silke Maes}
\email{silke.maes@kuleuven.be}

\author[0000-0003-4159-9964]{Silke Maes}
\affiliation{Institute of Astronomy, KU Leuven, Celestijnenlaan 200D, B-3001 Leuven, Belgium}

% \collaboration{20}{}

\author[0000-0001-5887-8498]{Frederik De Ceuster}
\affiliation{Institute of Astronomy, KU Leuven, Celestijnenlaan 200D, B-3001 Leuven, Belgium}

\author[0000-0001-9298-6265]{Marie Van de Sande}
\affiliation{Leiden Observatory, Leiden University, PO Box 9513, 2300 RA Leiden, The Netherlands}
\affiliation{School of Physics and Astronomy, University of Leeds, Leeds LS2 9JT, United Kingdom}

\author[0000-0002-5342-8612]{Leen Decin}
\affiliation{Institute of Astronomy, KU Leuven, Celestijnenlaan 200D, B-3001 Leuven, Belgium}
\affiliation{School of Chemistry, University of Leeds, Leeds LS2 9JT, United Kingdom}



\begin{abstract}
The chemistry of an astrophysical environment is closely coupled to its dynamics, the latter often found to be complex. Hence, to properly model these environments a 3D context is necessary. However, solving chemical kinetics within a 3D hydro simulation is computationally infeasible for a even a modest parameter study. In order to develop a feasible 3D hydro-chemical simulation, the classical chemical approach needs to be replaced by a faster alternative. We present \mace, a Machine learning Approach to Chemistry Emulation, as a proof-of-concept work on emulating chemistry in a dynamical environment.  Using the context of AGB outflows, we have developed an architecture that combines the use of an autoencoder (to reduce the dimensionality of the chemical network) and a set of latent ordinary differential equations (that are solved to perform the temporal evolution of the reduced features). Training this architecture with an integrated scheme makes it possible to successfully reproduce a full chemical pathway in a dynamical environment. \mace\ outperforms its classical analogue on average by a factor 26. Furthermore, its efficient implementation in PyTorch results in a sub-linear scaling with respect to the number of hydrodynamical simulation particles. %\mace\ is a promising tool to bridge the gap between theory and observations in the field of astrochemistry, and can be used to explore the chemical evolution in a 3D hydrodynamical environment.
\end{abstract}

\keywords{Astrochemistry (75) -- Computational methods (1965) -- Astronomy software (1855) -- Chemical reaction network models (2237) -- Asymptotic giant branch stars (2100) -- Stellar winds (1636)}


% \input{1-introduction.tex}
\section{Introduction} \label{sec:intro}
Astrochemistry, the study of chemistry in space, is a powerful tool. Combining observations of chemical species in astrophysical objects with theoretical predictions allows us to study the physical conditions, as well as to estimate its chemical composition and evolution. Astrochemistry labs are found in different environments ranging from dark clouds and protoplanetary disks to {different phases of interstellar medium (ISM), and cluster formation in} galaxies.
\\ \indent The astrophysical environment impacts its chemistry, and vice versa. For instance, cooling and heating processes as a result of chemical reactions will influence the dynamics. Hence, a hydrodynamics model needs to be coupled with a chemistry model{, apart from radiation,} in order to fully simulate an astrophysical environment. Moreover, this dynamics is often complex and therefore requires a 3-dimensional approach when modelling it. 3D hydrodynamical modelling is a notoriously computationally expensive process, both in a particle-based and grid-based approach. {It is often the case that} coupling such hydrodynamics with a classical chemical model in every time step makes it computationally infeasible to explore even a modest physical parameter space. {Various research groups have already made elaborate efforts in integrating (limited) chemistry in hydrodynamical simulations. To name a few, \cite{Glover2007I,Glover2007II, Walch2015}, and \cite{Hu2021} combine hydro and chemistry, amongst other processes, in the case of molecular clouds and the ISM, \cite{GRIFFINproject2020} incorporated both constituents in star formation simulations, \cite{Yoneda2016} and \cite{Young2021} did so for protoplanetary disk research (the latter doing chemistry in a post-processing step), and \cite{Richings_Schaye2016} for galaxy formation,. In this research, we take} an alternative pathway to produce feasible hydro-chemical simulations.% \citep{BoulangierI2019}. 
\\ \indent The astrophysical environment studied in this paper is the circumstellar envelope (CSE) of asymptotic giant branch (AGB) stars, i.e., evolved stars with an initial mass ranging between 0.8 and 8 solar masses. The CSE is created by the global stellar outflow launched at the surface of the AGB star (e.g., \citealp{Freytag2017}), believed to be due to the combination of surface pulsations facilitating dust formation \citep{Bowen1988,HofnerOlofsson}. Mass-loss rates are typically found to be between $10^{-8}$ and $10^{-4}$ $\Msolyr$, and terminal velocities are ranging from about 5 to 20 $\kms$ \citep{Knapp1998,Habing2004agbs.book,Ramstedt2009}. Thanks to the favourable physical conditions and large physical gradients in the outflow, the CSE hosts a rich chemistry; over 100 molecules and about 15 dust species have been detected so far \citep{Verhoelst2009,Water2011,Gail2013,Decin2021}. Recently, CSEs have been found to contain asymmetrical structures, such as spirals, arcs, disks, and bipolarity (e.g., \citealp{Mauron2006,Kervella2016,Li2016,Decin2021}). The most probable hypothesis is that the gravitational interaction with an unseen (sub)stellar companion shapes the outflow, causing the asymmetrical morphology \citep{Nordhaus2006,Decin2020,Gottlieb2022}. 3D hydrodynamical simulations affirm this hypothesis (e.g.\ \citealp{Theuns1993,Mastrodemos1999,KimTaam2012, ElMellah2020,Maes2021,Malfait2021}). While more complex physical processes are being implemented in hydrodynamical simulations step by step \citep{Maes2022}, such as radiation processes (e.g.\ \citealp{Chen2017,Esseldeurs2023}), no 3D hydro-\emph{chemical} simulation of these environments exists yet, which is necessary in order to start bridging the gap between theory and observations. 
\\ \indent Due to the expansion of CSEs, its chemical composition never reaches an equilibrium state. In these dynamical conditions, the evolution of the chemical abundances over time is described by a chemical kinetics model. More specifically, the change of number density of the chemical species in the desired chemical network is calculated by a set of non-linear, coupled ordinary differential equations (ODEs, e.g., \citealp{Millar2015}). Chemical kinetics, from a mathematical point of view, is known to be a stiff problem (e.g., \citealp{Wen2023}), due to (i) the short time scales associated with certain chemical variation in astrophysical environments and hence the need for small time steps in the solver, and (ii) the range in parameters that spans many orders of magnitude. Hence, computation is typically relatively slow for an extensive chemical network of a couple of hundred species. 
\\ \indent The classical way to work around the long computation times is to reduce the extensive chemical network, given an astrophysical environment, to its most important species and reactions only, since typically only a handful of chemical species dominate the overall chemistry. In order to do so, different routines exist (e.g.\ \citealp{Grassi2013}). For CSEs of AGB stars, this has been done, for example,  by \cite{BoulangierI2019}, for the chemistry in the inner part of the CSE with the aim to evaluate it simultaneously with hydrodynamics. In order to verify whether the reduced network properly represents the desired chemistry, one needs to perform an extensive parameter study over a large dataset and have error measures to quantify its performance.  %Whether or not a reaction is important, is determined by whether the chemical kinetics changes when leaving the reaction out of the chemical network.  
Reducing the chemical network has the advantage that it has a clear chemical interpretation: it leaves in the dominant chemical reactions that play a crucial role in the chemical kinetics, and removes reactions that do not contribute much. However, due to the non-linear nature of chemistry, secondary reactions, deemed unimportant at first sight, might be crucial in the end (e.g., as a result of chain reactions). {Therefore, it is beneficial to be able to include an extended chemical network for thoroughly studying the chemical destruction and formation pathways of species in a given astrophysical environment.}%Hence, one would never be able to do a full chemical analysis with a reduced chemical network.
\\ \indent Another way to speed up the computation time is by building a surrogate model, i.e., a model that is able to emulate the chemistry with a (severely) reduced computation time. In recent years, many different dimensionality reduction and function approximation techniques have been developed, commonly categorised within the field of Machine Learning (ML), that show promising prospects in the field of (astro)chemistry \citep{Wen2023}. %To emulate chemical evolution, mainly supervised learning is applicable, i.e.\ training a surrogate model on labelled data (namely, the models made with classical chemical kinetics). 
In the framework of chemistry in the interstellar medium, \cite{Grassi2011} were the first to emulate chemical kinetics with a simple neural network (NN). Later, \cite{deMijola2019} have shown that an immense speed up, up to a factor of $10^5$, can be gained with this approach. \cite{Holdship2021} and \cite{Grassi2022} both used an autoencoder architecture \citep{Kramer1992} to first reduce the dimensionality of their chemical network and subsequently evolve the chemical kinetics on this reduced network. The way they evolve the chemistry however differs; while in \cite{Holdship2021} another NN handles this, \cite{Grassi2022} train the autoecoder to be interpretable still in the reduced chemical space, and solve the reduced chemistry with a interpretable set of ODEs. \cite{Sulzer2023} adopt a similar method as the latter, but step away from the physical interpretation of the reduced chemical network, and use a linear function to evolve the reduced features. Additionally, other ML strategies have been used to speed up the computation time of chemical models. \cite{Branca2022} opted for physics informed neural networks (PINN, \citealp{Raissi2019}), and in a follow-up work \citep{Branca2024} they explored the use of a deep neural operator (DeepONet, \citealp{Lu2021}), while \cite{Palud2023} make use of NNs as a regression technique. 
\\ \indent However, the crucial caveat in these works is that the time evolution of the chemistry is performed in a \emph{static} physical environment. The objective of this work is to develop a scheme that is able to emulate the chemical evolution in a \emph{dynamical} physical environment, namely with changing physical parameters over time. 
\\ \indent As such, we introduce \mace\ -- a \textit{Machine learning Approach to Chemistry Emulation}. With \mace\ we step away from the classical way of calculating chemical evolution in a dynamic environment. We follow a similar approach as \cite{Sulzer2023}, and map the chemical kinetics problem to a reduced dynamical system that does no longer necessarily have a physical or chemical interpretation. Hence, we deliberately sacrifice the interpretability of our model to allow for more freedom to optimise the trade-off between the accuracy and the computational speed of the resulting surrogate model. 
\\ \indent This paper is organised as follows. Sect.\ \ref{sect:methods} describes the different methods to calculate chemical abundances. In Sect.\ \ref{sect:mace}, we introduce the architecture of our emulator \mace\ and in Sect.\ \ref{sect:training} its training methodology is presented in detail. Sect.\ \ref{sect:results} encompasses the results of different trained \mace\ models, which are further discussed in Sect.\ \ref{sect:discussion}. In the latter we also suggest some future improvements. Finally, we conclude in Sect.\ \ref{sect:conclusion}.

\section{Methodology}\label{sect:methods}
Before going into details about \mace, we first introduce the methodology of classical chemical kinetics, how this translates to emulation, and describe its application specifically on AGB circumstellar envelopes.

\subsection{Classical chemical kinetics}\label{sect:classical}
Classically, given a physical state (commonly determined by density, temperature, and radiation field), the evolution over time of the abundances of the chemical species is described by a chemical kinetics model. This model describes the rate of change of the number density $n$ by solving a set of coupled, non-linear ordinary differential equations (ODEs). For a certain chemical species $i$, such a rate equation is given by
\begin{equation}\label{eq:ODE_gen}
    \frac{{\rm d}n_i}{{\rm d}t} = \sum_{j\in F_i}\left(k_j \prod_{r\in R_j}n_r\right)-\sum_{j\in D_i}\left(k_j \prod_{r\in R_j}n_r\right).
\end{equation}
It states the balance between the formation reactions $F_i$ (first term), and destruction reactions $D_i$ (second term) of species $i$ with reactant $r$ from the set of reactants $R_j$, where $k_j$ is the rate coefficient of reaction $j$. The rate $\frac{{\rm d}n_i}{{\rm d}t}$ has units ${\rm cm^{-3}\, s^{-1}}$. %Its given in units of s$^{-1}$, including a spatial component depending on the number of reactants involved. 
If the chemical network only contains one-body and two-body reactions, the rate equation for species $i$ reduces to
\begin{equation}\label{eq:ODE}
    \frac{{\rm d}n_i}{{\rm d}t} = \sum_{j,l}k_{jl}n_j n_l + \sum_{m}k_m n_m-n_i\left(\sum_r k_{ir}n_r+\sum_s k_s\right), %\quad [{\rm cm^{-3}\, s^{-1}}].
\end{equation}
where now the indices $j,l,m,r,s$ run over all other chemical species in the network. We can write this more succinctly in matrix notation as 
\begin{equation}\label{eq:ODE_matrix}
    \frac{{\rm d}n_\alpha}{{\rm d}t} = A_{\alpha\beta} \, n_\beta + B_{\alpha\beta\gamma}\, n_\beta \, n_\gamma,
\end{equation}
where we sum over repeated indices. $A_{\alpha\beta}$ is a matrix containing the one-body reaction coefficients $k_m$ and $k_s$ from Eq.\ (\ref{eq:ODE}), in units of s$^{-1}$, and $B_{\alpha\beta\gamma}$ is a tensor with the two-body reaction rates $k_{jl}$ and $k_{ir}$ as elements, in units of cm$^3$\,s$^{-1}$. Henceforth, we will indicate matrices containing the coefficients in bold to clearly indicate their multi-dimensional nature. $\boldsymbol{A}$ and $\boldsymbol{B}$ are sparse, since chemical species only react with a limited number of other species. %Most often, the set of coupled ODEs given in Eq.\ (\ref{eq:ODE_matrix}) is stiff, since certain chemical abundance change quickly over time and the range of number densities and rate coefficients involved, differ many orders of magnitude.  

\subsection{Emulating chemical kinetics}\label{sect:emulation}
We build a surrogate model that is able to emulate the chemical evolution given by Eq.\ (\ref{eq:ODE_matrix}). More specifically, we map the chemical kinetics system to a reduced dynamical system that can be solved more efficiently. The emulator takes the initial set of abundances $\n_0$ as input, together with the physical input $\p$. Subsequently, the emulator predicts the evolution of the chemical species over a given time $t$, i.e., the set of abundances $\hat{\n}(t)$, where the $\hat{\cdot}$\,-symbol is used for predictions made by the emulator. In Sect.\ \ref{sect:mace}, we elaborate on its architecture.

\begin{figure*}
    \centering
    \includegraphics[width=0.8\textwidth]{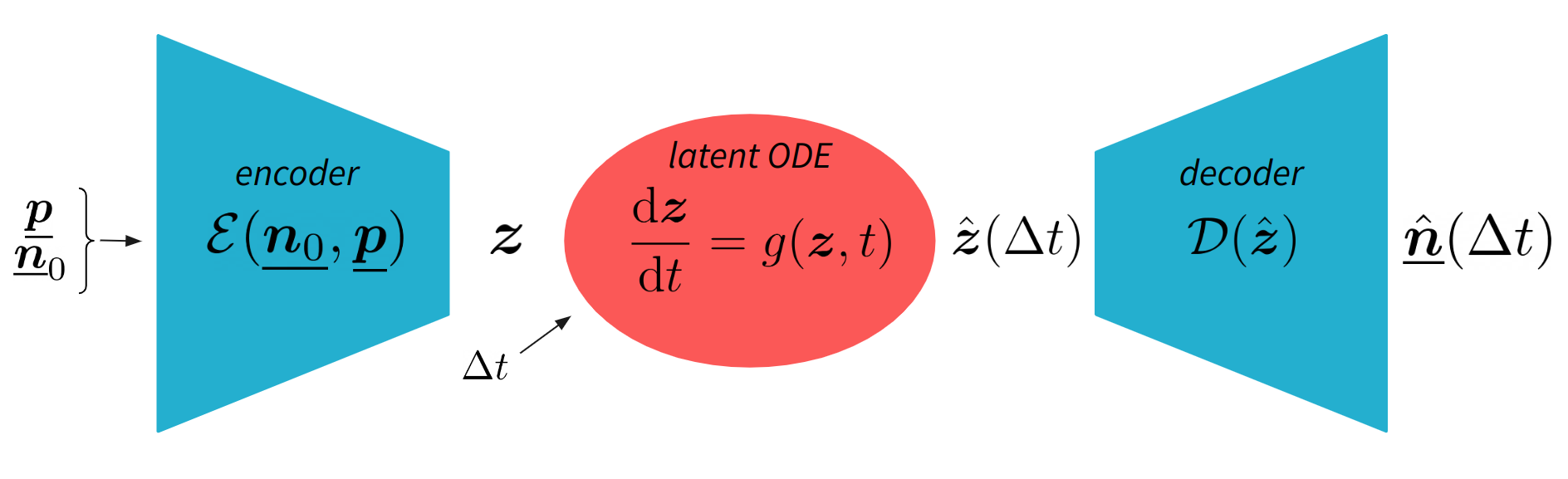}

    \caption{Schematic representation of the architecture of \mace. The autoencoder (encoder $\E$ and decoder $\D$) is given in blue. The latent ODE $g$ is given in red. The flow of the surrogate model is from left to right. The initial abundances $\underline{\n}_0$ and physical parameters $\underline{\p}$ are concatenated and fed into the encoder, resulting in the latent representation $\z$. This latent representation is fed into the latent ODE function $g$, together with a timestep $\Delta t$. Solving this differential equation results in the predicted evolution in latent space $\hat{\z}(\Delta t)$. This predicted latent vector is then fed into the decoder, resulting in the predicted abundances $\underline{\hat{\n}}(\Delta t)$.}
    \label{fig:arch}
\end{figure*}

\subsection{Application: AGB circumstellar envelope}\label{sect:testcase}
A 1-dimensional representation of an AGB star's circumstellar envelope is used as the astrochemical environment for the development of \mace. Since this environment is already well studied and understood (e.g., \citealp{Millar2000,Li2016,VandeSande2019,Maes2023}), \mace\ can  be benchmarked against this well-studied case. To generate training data, we use a version of the publicly available CSE model of the UMIST Database for Astrochemistry (UDfA)\footnote{The model can be found on GitHub: \url{https://github.com/MarieVdS/rate22_cse_code} \citep{Millar2024}}, which calculates the chemical abundances as a function of distance from the star (using a modified form of Eq.\ \ref{eq:ODE}). We use an adaptation of the chemical network \textsc{rate}12\footnote{\label{footnote:UDfA} The network can be found online: \url{http://udfa.ajmarkwick.net/index.php?mode=downloads}.}. The network consists of gas-phase chemistry only, involving 468 different chemical species (including electrons as a separate species), that can interact via 6180 reactions. 
{Different types of reactions are included, such as two-body reactions between neutral and ionised species, photodissociation, and cosmic-ray induced reactions (for details, see \citealp{McElroy2013}).}
We will further refer to this model as the `\csone' model. 
\\ \indent The \csone\ model assumes a power law for the density $\rho$ and the temperature $T$, as a function of radius:
\begin{eqnarray}
    \rho(r) &=& \frac{\Mdot}{4\pi r^2 \vexp}, \label{eq:density}\\
    T(r) &=& \Tstar \left(\frac{r}{\Rstar}\right)^{-\varepsilon}. \label{eq:temp}
\end{eqnarray}
The mass-loss rate $\Mdot$ and the expansion velocity $\vexp$ set the density, whereas the exponent $\eps$ sets the steepness of the temperature gradient. Here, $\Tstar$ is the temperature of the AGB star at its surface $\Rstar$. H$_2$ is assumed to be fully self-shielding. Further details about the model can be found in, e.g., \cite{Millar2000}, \cite{CM2009}, \cite{McElroy2013}, and \cite{VdS2018}. In Sect.\ \ref{sect:dataset}, we elaborate on the specific dataset used for the training.

\section{Emulator architecture}\label{sect:mace}
The architecture of the \mace\ emulator is chosen with two goals in mind: (i) we aim to take into account the full chemical network of \textsc{rate12}, as such we want to reduce its dimensionality, (ii) we aim for a flexible emulator that is not restricted to a specific evolution time step. That is, the emulator should be able to accurately predict the evolution pathway of the chemical species over time without errors growing too much throughout the evolution. Hence, the architecture of \mace\ consists of two main parts: (i) an autoencoder for the dimensionality reduction, and (ii) a trainable ODE as substitute for the chemical evolution, on both we elaborate later in this section. Schematically, we can write \mace\ as the following function:
\begin{equation}\label{eq:mace}
    \underline{\hat{\n}}(t) = \D\Big( G \big( \E (\underline{\n}, \underline{\p}),t \big) \Big),
\end{equation}
where the underlined symbols indicate that that parameter is preprocessed and not used in its ``raw'' form (see further in Eq.\ \ref{eq:transf} in Sect.\ \ref{sect:dataset}).
$\E$ and $\D$, the encoder and decoder respectively, constitute the autoencoder, while the function $G$ describes the evolution in latent space such that $\z(\Delta t) = G(\z, \Delta t)=\int_0^{\Delta t} g(\z){\rm d}t$, with $\z$ the latent space variables. A schematic visualisation of the architecture is given in Fig.\ \ref{fig:arch}. \mace\ is built in Python using the PyTorch framework \citep{pytorch} and is publicly available on GitHub:\ \url{https://github.com/silkemaes/MACE}.

\subsection{Autoencoder}\label{sect:ae}
Autoencoders are a widely used tool to reduce the dimensionality of a certain set of data and typically consist of an encoder and a decoder, $\mathcal{E}$ and $\mathcal{D}$ in Eq.\ \eqref{eq:mace}, respectively. They are a type of neural network architecture that are trained to reproduce their input as output. The encoder takes the input and maps it to a latent space, the lower-dimensional representation of the high-dimensional input. The decoder then maps the latent space back to the original input space. The latent space is thus a compressed representation of the input \citep{Kramer1992}. 
\\ \indent We construct a purely mathematical representation of the chemical network in latent space. Since in a chemical reaction network of 468 chemical species only a couple of a dozen of species are dominating the chemical pathways, it is expected that the dimensionality of the chemical network can be reduced by an order of magnitude at least, without losing important information \citep{Holdship2021,Grassi2022,Sulzer2023}. 
\\ \indent For \mace, the encoder $\mathcal{E}$ takes as input the set of chemical abundances $\underline{\n}$, concatenated with the physical parameters $\underline{\p}$, resulting in $(468+4)$ nodes in the input layer. Mathematically, it can be written as
\begin{equation}\label{eq:encoder}
    \z = \E(\underline{\n}, \underline{\p}), 
\end{equation}
where $\z$ represents the parameters in latent space. The input layer is followed by two hidden layers and an output layer. The output layer has a number of nodes equal to the dimensionality of the latent space $d$, which is varied in the training stage (Sect.\ \ref{sect:hyperparames}). The number of nodes in each hidden layer decreases and is chosen to equal a power of 2 in order to optimise computational resources, namely 256 and 64. The activation function in every layer, except for the output layer, is a leaky rectifier (leaky ReLU) with slope 0.2 \citep{Maas2013}. For the output layer, a hyperbolic tangent ($\tanh$) is used, in order to map the input to values within a range $]-1,1[$ in latent space\footnote{We want to restrict the range in values in the latent space to better control the latent dynamics (see Sect.\ \ref{sect:latdyn}).}. 
\begin{table}
    \begin{center}
        \caption{Number of nodes in the different layers of the autoencoder (encoder + decoder). The parameter $d$ indicates the number of dimension in latent space.}
        \begin{tabular}{   r     c c c c        }
            \hline \hline \\[-2ex]
                     & input         & hidden 1    & hidden 2   & output         \\ \hline
       encoder    & $468+4$         & 256         & 64         & $d$  \\
       decoder    & $d$ & 64          & 256        & 468            \\ \hline						
       \end{tabular}
        \label{tab:ae}
    \end{center}
    {\footnotesize \textbf{Notes.} For all layers, except the output layer of the encoder, a leaky ReLU function with slope 0.2 is used as activation function. For the output layer of the encoder, a $\tanh$ is used.}
\end{table}
\\ \indent The decoder $\mathcal{D}$ has the same architecture as the encoder, but with the layers of the encoder reversed. In the decoder only leaky ReLUs with a slope of 0.2 are used as activation function. The number of nodes in the output layer of the decoder matches the number of chemical species. Mathematically, the decoder is given by 
\begin{equation}\label{eq:decoder}
    \underline{\hat{\n}} = \D(\z).
\end{equation}
\\ \indent To train the autoencoder, we aim for the following expression to be satisfied:
\begin{equation}\label{eq:ae}
    \underline{\n} = \D(\E(\underline{\n},\underline{\p})).
\end{equation}
An overview of the layers and number of nodes in the autoencoder can be found in Table \ref{tab:ae}.

\subsection{Latent ODE}\label{sect:latentODE}
Inspired by the mathematical form of the chemical rate equations, Eq.\ (\ref{eq:ODE}), we opted for the latent set of ODEs to be of the same form and include a constant term, akin to \cite{Sulzer2023}. The latent ODE is given by
\begin{align}
    \frac{{\rm d}z_\alpha}{{\rm d}t}&= g(z_\alpha,t) \nonumber \\
    &= \mathcal{C}_\alpha + \mathcal{A}_{\alpha\beta}z_\beta + \mathcal{B}_{\alpha\beta\gamma}z_\beta z_\gamma,\label{eq:latentODE}
\end{align}
where $\z$ is the encoded latent representation of ($\underline{\n}$,$\underline{\p}$), Eq.\ \eqref{eq:encoder}, and $\C$, $\A$, and $\B$ are constant tensors with dimensions matching the operation. The elements in these tensors are trainable parameters that are optimised during training. We solve the latent ODE (Eq.\ \ref{eq:latentODE}) using the PyTorch library \emph{torchode} \citep{lienen2022torchode}, allowing to train it at the same time due to its gradient tracking. Explicitly solving the latent ODE adds complex dynamics to the emulator, contrary to \cite{Sulzer2023}'s approach, which makes it better grasp the complexity of the chemical kinetics model.

\subsection{Latent dynamics}\label{sect:latdyn}
%Finding a non-exponentially growing solution for the latent ODE (Eq.\ \ref{eq:latentODE}) will often not be possible for an arbitrary time $t$, since most often the eigenvalues of this equation will be positive for a random input. Since here we are working in latent space, time $t$ has lost its physical interpretation, and can therefore be scaled at will. {\color{red} @Frederik, dit is denk ik hoe ik het versta, voel u vrij om dit wat aan te commenten en het wiskundig correct(er) te maken!}
The dynamics of a general set of coupled, non-linear ODEs, such as the one given by Eq.\ \eqref{eq:latentODE}, can be very chaotic without any constraints on the values of the coefficients. As such, the overall behaviour of the system is difficult to control. For instance, the evolution of a system as in Eq.\ \eqref{eq:latentODE}, initialised with random coefficients, will typically make one or more components diverge. This will complicate the training process; on the one hand we want to explore the space of latent dynamics as much as possible, but on the other hand, the typical divergence of the system will regularly cause problems in any attempt to numerically solve it. One way to resolve this is by constraining the latent dynamics only to ``well-behaved'' systems of ODEs that do not cause any divergences over the relevant time span. Although there is plenty of mathematical literature on the behaviour of dynamical systems that could help constrain the coefficients of our latent dynamics (e.g., \citealp{Strogatz2000}), we opted for a more empirical approach.
\\ \indent Our approach is based on a simple observation: for any dynamical system of the form Eq.\ \eqref{eq:latentODE} with random (but finite) coefficients and initial conditions $\z_d(0)$ of dimension $d$, the width of the distribution of evolved latent variables $\z_d(t)$ can be controlled by the time $t$. In latent space, the time variable $t$ no longer has any physical meaning, so we can rescale it to control the latent dynamics. The appropriate rescaling of the time variable will depend on the distributions of the ODE coefficients and the initial conditions, but most importantly on the dimension $d$ of the dynamical system.
\begin{figure}
    \centering
    \includegraphics[width=0.475\textwidth]{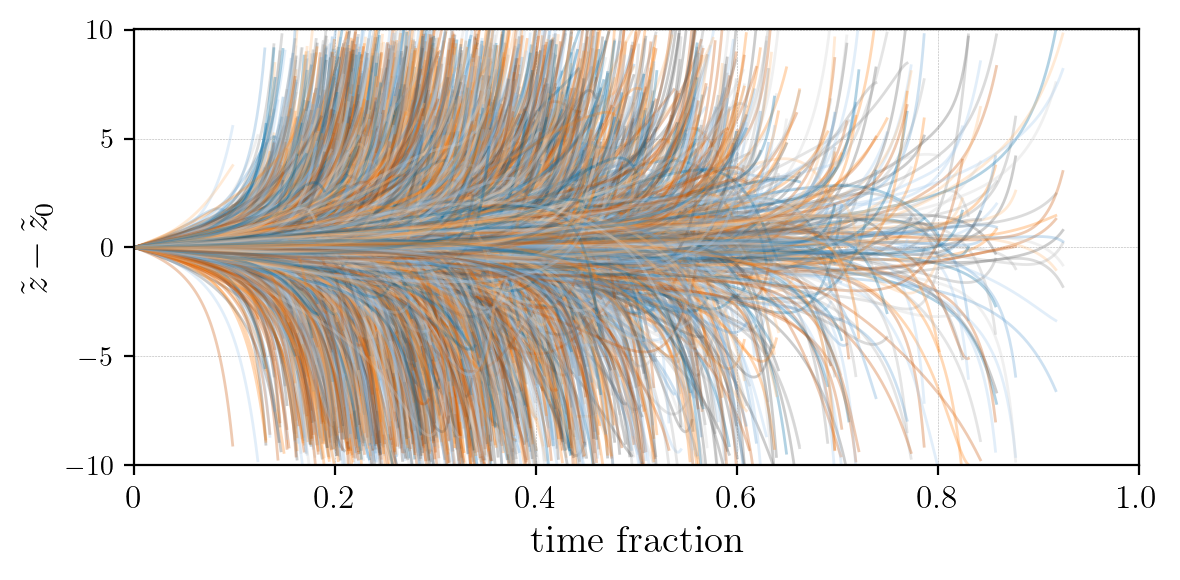}
    \caption{Example of result of an empirical latent dynamics test. The evolution as a function of time (given in time fraction) is given for 1\,000 randomly generated $\boldsymbol{\Tilde{z}}_{8}$, following Eq.\ \eqref{eq:latentODE}.}
    \label{fig:z_dyntest}
\end{figure}
\begin{figure}
    \centering
    \includegraphics[width=0.475\textwidth]{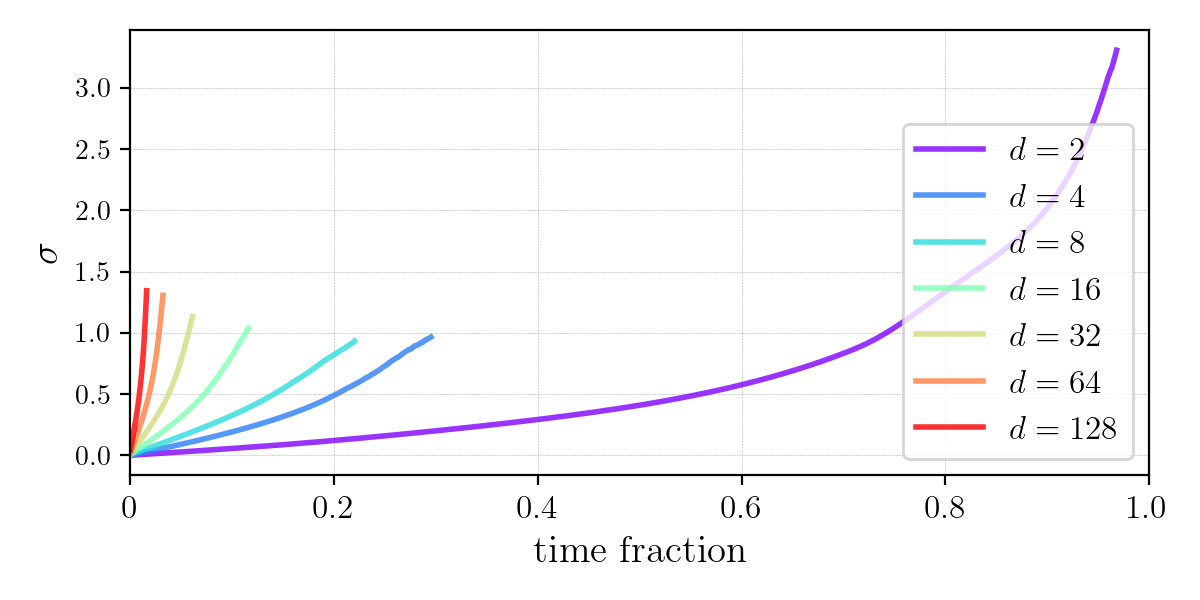}
    \caption{Standard deviations $\sigma$ for the spread in $\Tilde{\z}_{d}$ for the empirical tests of the latent dynamics (Eq.\ \ref{eq:latentODE}) for different dimensionality $d$, as a function of time fraction. The cutoff of each curve happens when, at that fraction in time, less than 95\% of the empirical solutions are no longer bound and thus diverge to infinity. }
    \label{fig:all_dyntest}
\end{figure}
\begin{figure}
    \centering
    \includegraphics[width=0.475\textwidth]{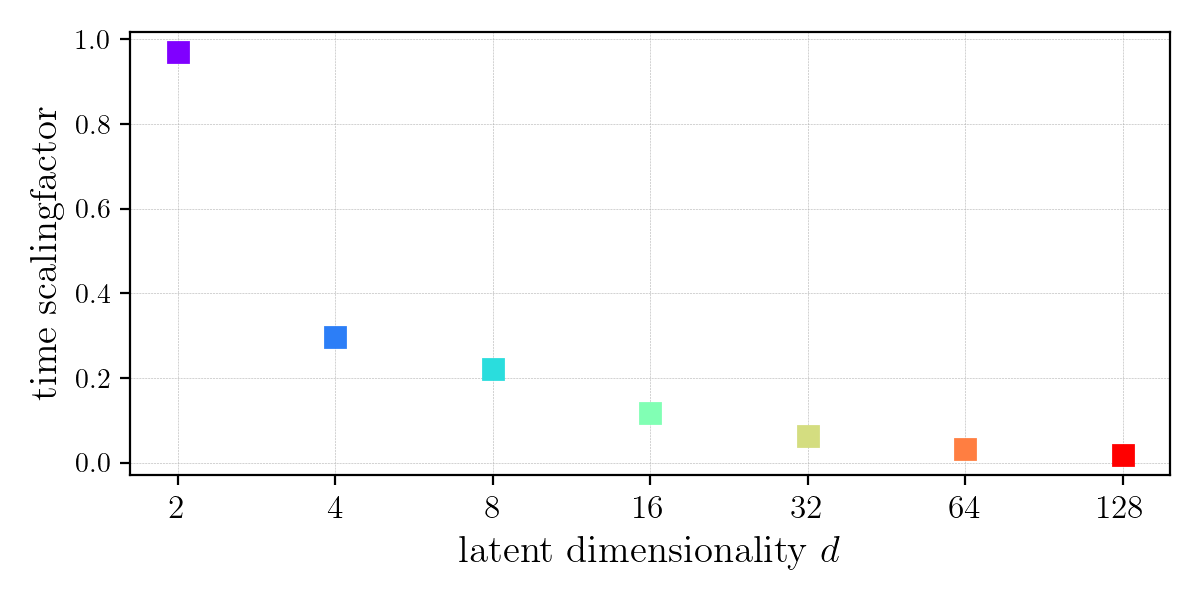}
    \caption{Resulting time rescaling per latent dimensionality $d$ from empirical latent dynamics test. The colours correspond to the colours in Fig.\ \ref{fig:all_dyntest}.}
    \label{fig:timefract_dyntest}
\end{figure}
\\ \indent Determining the optimal rescaling is done with an empirical strategy using a numerical experiment. We produce 10\,000 random initial values for the latent space vector, $\boldsymbol{\Tilde{z}}_d(0)$ (the $\boldsymbol{\Tilde{.}}\,$-symbol indicating the empirical test), uniformly distributed between -1 and 1 to match the possible outcome of the encoder (see Sect.\ \ref{sect:ae}), as a function of different latent dimensionality $d$. Given the tensors $\boldsymbol{\Tilde{\mathcal{C}}}$, $\boldsymbol{\Tilde{\mathcal{A}}}$, and $\boldsymbol{\Tilde{\mathcal{B}}}$ with random, standard normally distributed values, Eq.\ (\ref{eq:latentODE}) is solved for each $\boldsymbol{\Tilde{z}}_d$ using \emph{torchode} \citep{lienen2022torchode}. An example of the outcome of such an empirical test for $d=8$, i.e., the evolution of $\boldsymbol{\Tilde{z}}_{8}$ over a normalised time, is shown in Fig.\ \ref{fig:z_dyntest}.
\\ \indent In order to determine the appropriate rescaling of the time parameter, the following methodology is used: (i) We require that a cutoff of 95\% of empirical tests should still have a bound solution after a time $t_{\rm cutoff}$. This time $t_{\rm cutoff}$, normalised to the full evolution time, will serve as the scaling factor. (ii) We perform empirical tests for dimensionality $d\in[2,4,8,16,32,64,128]$. (iii) We look for the time $t_{\rm cutoff}$ by calculating the standard deviation $\sigma$ as a function of evolution time. This $\sigma$ serves as a measure of the range of the dynamics of an empirical latent system. The $\sigma$'s are only computed when at least 95\% of the models still has a bound solution, giving us $t_{\rm cutoff}$. Fig.\ \ref{fig:all_dyntest} shows the $\sigma$'s for different latent dimensionality. The resulting time scale factors, after normalising the $t_{\rm cutoff}$ are shown in Fig.\ \ref{fig:timefract_dyntest}. For increasing dimensionality, the scaling fraction decreases, since a higher latent dimensionality increases the chance of a divergence.

\newpage
\section{Training}\label{sect:training}
Training an emulator is a non-trivial optimisation problem involving numerous free parameters. In this section, we elaborate on the parameter space, the data used for training, the different loss functions that are involved in the optimisation, and distinguish two training schemes.

\subsection{Parameter space \& dataset}\label{sect:dataset}
To be efficient with data and computational resources, we reuse the grid of chemical models from the sensitivity analysis performed by \cite{Maes2023} as training data for \mace. The grid was constructed to span a broad range of CSE parameters found of AGB stars, based on observations. These \csone\ models were generated using the CSE model introduced in Sect.\ \ref{sect:testcase} and consists of 18\,000 models of the chemistry in carbon-rich CSEs for varying densities and temperatures. The set of initial abundances (i.e., parent species) specific to carbon-rich AGB outflows can be found in Table \ref{tab:parents}.
\begin{table}
    \begin{center}
        \caption{Parent species of the carbon-rich AGB outflows, and their initial abundances relative to H$_2$.}
        \begin{tabular}{   l   r        }
            \hline \hline \\[-2ex]
            {Species} & {Abundance }   \\ \hline
            He          & $0.17$   \\ 
            CO          & $8.00\times10^{-4}$  \\ 
            C$_2$H$_2$  & $4.38\times10^{-5}$   \\ 
            HCN         & $4.09\times10^{-5}$   \\ 
            N$_2$       & $4.00\times10^{-5}$   \\ 
            SiC$_2$     & $1.87\times10^{-5}$   \\ 
            CS          & $1.06\times10^{-5}$   \\ 
            SiS         & $5.98\times10^{-6}$   \\ 
            SiO         & $5.02\times10^{-6}$   \\ 
            CH$_4$      & $3.50\times10^{-6}$   \\ 
            H$_2$O      & $2.55\times10^{-6}$  \\ 
            HCl         & $3.25\times10^{-7}$   \\ 
            C$_2$H$_4$  & $6.85\times10^{-8}$   \\ 
            NH$_3$      & $6.00\times10^{-8}$   \\ 
            HCP         & $2.50\times10^{-8}$   \\ 
            HF          & $1.70\times10^{-8}$   \\ 
            H$_2$S      & $4.00\times10^{-9}$   \\ 
             \hline						
        \end{tabular}
        \label{tab:parents}
    \end{center}
    {\footnotesize \textbf{Notes.} Abundances taken from \cite{Agundez2020}. When a range was given there, the linear average is used.}
\end{table}
\\ \indent Each of the models spans an outflow radius from $10^{14}$ to $10^{18}\,cm$ measured from the centre of the star, with a radial resolution of 134 steps. The ranges of the different parameters that were used to build the grid of \csone\ models are given in Table \ref{tab:grid}. A visualisation of the density and temperature space can be found in Appendix \ref{sect:physpar_app}. Since the models assume a constant outflow velocity, we can relate the distance scale to time scale, using that $t = r/\vexp$. Hence, this results in a maximal chemical evolution time ranging from $0.4\times10^{12}$\,s to $4\times10^{12}$\,s for the range of expansion velocities considered. 
\begin{table}
    \begin{center}
        \caption{Physical parameters and their ranges of the grid of chemical models from \cite{Maes2023}. The density is determined by $\Mdot$ and $\vexp$ according to Eq.\ (\ref{eq:density}) for the combinations given in Fig.\ \ref{fig:grid_dens}, the temperature profile is given by the combination of $\Tstar$ and $\eps$ through Eq.\ (\ref{eq:temp}). The stellar radius, $\Rstar$, inner radius, $R_{\rm inner}$, and outer radius, $R_{\rm outer}$, are kept constant. For the purpose of this research, the radii are converted to time, using the expansion velocity of the corresponding simulations. Adapted from \cite{Maes2023}.}
        \begin{tabular}{   l l   c   c    }
            \hline \hline \\[-2ex]
            Parameter & & Range/Value & Stepsize   \\ \hline
            $\Mdot$ &[$\Msolyr$] &$1\times10^{-8}$ -- $5\times 10^{-5}$ & (*) \\
            $\vexp$ &[$\kms$] & 2.5 -- 25 & 2.5	 \\ 
            $\Tstar$ &[K] & 2000 -- 3000 & 50 \\ 
            $\varepsilon$& / & 0.3 -- 1.0 & 0.05  \\ \hline
            $\Rstar$ & [cm] & $2\times 10^{13}$ & / \\
            $R_{\rm inner}$ & [cm] & $10^{14}$ & /  \\ 
            $R_{\rm outer}$ & [cm] & $10^{18}$ & / 
            \\ \hline
        \end{tabular}
        \label{tab:grid}
    \end{center}
    {\footnotesize \textbf{Note.} (*) For $\Mdot$ we used the values $1\times10^{-p}$, $2\times10^{-p}$ and $5\times10^{-p}$, with $p\in[5,6,7,8]$, see also Fig.\ \ref{fig:grid_dens}.  
    }
\end{table}
\\ \indent The chemical evolution in the grid of models is determined mainly by four physical parameters
\begin{equation}\label{eq:physpar}
    \p = (\rho, T, \xi, A_V),
\end{equation}
where $\rho$ is the density, $T$ the temperature, and the two latter parameters denote the radiation field: $\xi$ sets the mean interstellar radiation field \citep{Draine1978}, normalised over a 3D sphere, and $A_V$ is the outward dust extinction in the visible part of the spectrum, thus can be seen as an optical depth. {As such, only the rate of photodissociation reactions is varied and not the cosmic-ray rate, since in smooth outflows it is found cosmic rays to not play a significant role (e.g., \citealp{VdSMillar2019})}. Hence, the vector $\p$ in Eq.\ \eqref{eq:physpar} is the input for the physical parameters in the \mace\ architecture (Eq.\ \ref{eq:mace}). 
\\ \indent To build the training dataset from the grid of chemistry models from \cite{Maes2023}, the density $\rho$ and temperature $T$ are calculated as given by Eqs.\ (\ref{eq:density}) and (\ref{eq:temp}), respectively. The two remaining parameters $\xi$ and $\AV$ depend indirectly on $\rho$. $\AV$ is calculated as 
\begin{equation}\label{eq:AV}
    A_{\rm V} = \frac{A_{\rm UV}}{[A_{\rm UV}/A_V]},
\end{equation}
where $A_{\rm UV}$ is the extinction in the UV part of the electromagnetic spectrum and the ratio $[A_{\rm UV}/A_{\rm V}]=4.65$ \citep{Nejad1984}. When we assume the extinction to be the same as that of the interstellar medium of $1.87\times10^{21} \, {\rm atoms\ cm^{-2}\ mag^{-1}}$ \citep{Cardelli1989}, then $A_{\rm UV}$ is given by
\begin{equation}\label{eq:AUV}
    A_{\rm UV} = [A_{\rm UV}/A_{\rm V}] \frac{N_{{\rm H}_2}}{1.87\times10^{21} },
\end{equation}
where $N_{{\rm H}_2}=\rho_{{\rm H}_2}\times r$ is the column density of H$_2$. The parameter $\xi$ is implemented as described by \cite{JM1981}, calculating its value by taking the mean over a 3D sphere and depends on $A_{\rm UV}$ (Eq.\ \ref{eq:AUV}). Fig.\ \ref{fig:physpar_expl} in Appendix \ref{sect:physpar_app} shows the physical parameters $\p$ as a function of radius and time for an example \csone\ model, with $\Mdot=1\times10^{-6}\,\Msolyr$, $\vexp=15\,\kms$, $\Tstar=2500\,$K, and $\eps=0.6$, resembling an average AGB outflow.
\\ \indent The data at separate timesteps in one \csone\ model are considered as separate training samples. Hence, each training sample has a constant set of physical parameters $\p$. We will refer to this training data as `\cszero' samples. Since every \csone\ model has 134 steps in time, we have $18\,000 \times 134 = 2.41\times10^{6}$ \cszero\ samples in total available for training, validating, and testing \mace.
%During the training, we use a split of $70\%-30\%$ for training and validation data, respectively.
\\ \indent The physical parameters of the \cszero\ samples, as well as the abundances, span orders of magnitude. Generally, this makes it more difficult to accurately train an architecture. Before we feed the samples to \mace, we perform the following transformation for the physical parameters to bring the input values closer together in terms of order of magnitude.
\begin{equation}\label{eq:transf}
    \underline{\p} \equiv \mathcal{F}\left(\log_{10}(\p)\right),
\end{equation}
where $\mathcal{F}$ is the min-max rescaling function given by 
\begin{equation}
    \mathcal{F}(\x) = \frac{\x-\min(\x)}{\max(\x)-\min(\x)},
\end{equation}
which normalises the input and brings its values in the range $[0,1]$. For the input abundances, we first clip them to $10^{-20}$, since this was the absolute tolerance of the ODE solver of the \csone\ models. We then perform the same transformation as for the physical parameters, given in Eq.\ (\ref{eq:transf}).

\subsection{Loss functions}\label{sect:loss}
\begin{figure}
    \centering
    \includegraphics[width=0.475\textwidth]{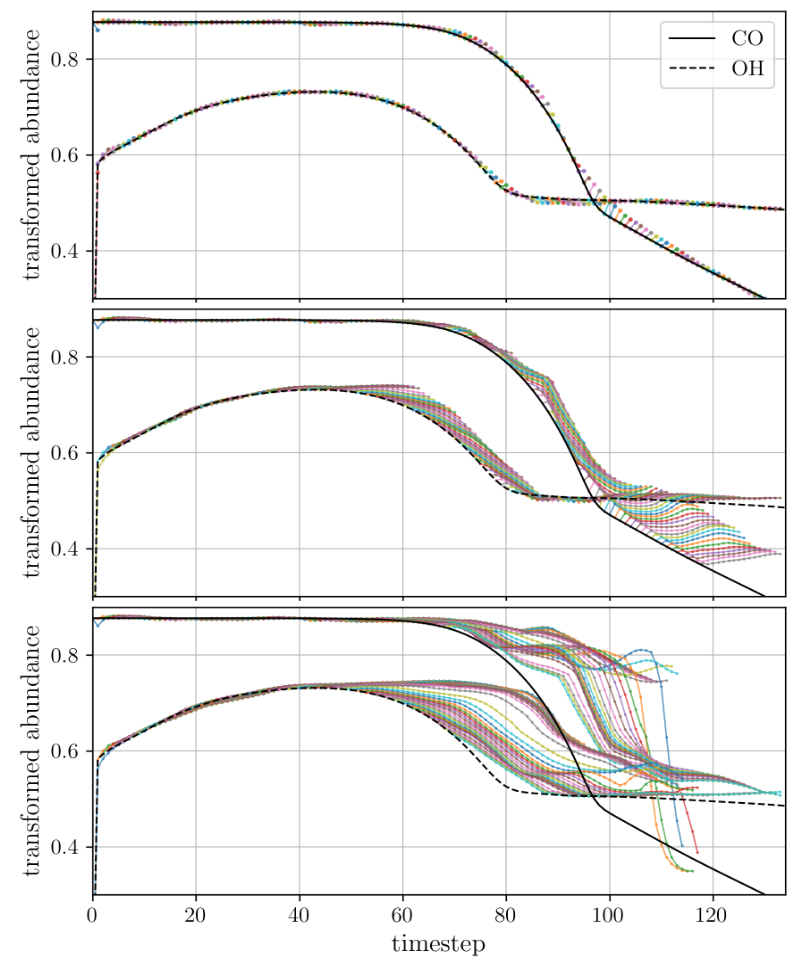}
    \caption{Visualisation of the integrated training scheme for different amount of timesteps $m$; \underline{Upper panel:} $m=1$ (hence, this reduces to the local training). \underline{Middle panel:} $m=16$, \underline{Bottom panel:} $m=64$. During the training, the predicted chemical paths (coloured curves) are matched to the real path (black curves). An example is given for the abundance profiles of CO (full curve) and OH (dashed curve). The \csone\ model used here has the following parameters: $\Mdot=1\times10^{-8}\,\Msolyr$, $\vexp=10\,\kms$, $\Tstar=2600\,$K, and $\eps=0.7$. }
    \label{fig:vis_loss}
\end{figure}

In order to train \mace, we define three localised losses and one integrated loss to express the desired behaviour of \mace\ that we want to optimise for.
\begin{itemize}
    \item The \emph{local absolute loss} (ABS) is defined as
    \begin{equation}\label{eq:ABS}
        \boldsymbol{L}_{\rm ABS} = |\underline{\n}(t)-\underline{\hat{\n}}(t)|,
    \end{equation}
    where \underline{$\n$} and \underline{$\hat{\n}$} are the real and predicted abundance, both transformed according to Eq.\ \eqref{eq:transf}, respectively. The aim of the absolute loss is to enforce \mace\ to match the predicted values of the abundances with the real ones. Since the abundances of the different species are rescaled to a range of $[0,1]$, the absolute loss will treat all species with a more equal importance compared to the unscaled abundances.
    \item The \emph{local gradient loss} (GRD) is calculated as
    \begin{equation}\label{eq:GRD}
        \boldsymbol{L}_{\rm GRD} = \Big|\frac{{\rm d}\underline{\n}}{{\rm d}t}-\frac{{\rm d}\underline{\hat{\n}}}{{\rm d}t}\Big|,
    \end{equation}
    which aims to enforce that the evolution of the abundances matches with the classical model, rather than only the instantaneous values.
    \item The \emph{identity loss} (IDN) is defined as
    \begin{equation}\label{eq:IDN}
        \boldsymbol{L}_{\rm IDN} = |\underline{\n}-\D\big(\E(\underline{\n},\underline{\p})\big)|,
    \end{equation}
    with $\E$ the encoder and $\D$ the decoder, respectively. It is used to enforce that the autoencoder reproduces the input as its output, and thus minimises the loss of information during the compression of the real chemical space to the latent space. Note here that the identity loss only acts on the real abundances $\underline{\n}$ at a certain time (and their un-evolved, autoencoded state), and not on the predicted abundances $\underline{\hat{\n}}$.
    \item We define a \emph{time-integrated absolute loss} (iABS), which enforces to match a chemical \emph{pathway} instead of only the next step in time. More specifically, this loss compares the predicted path (i.e., evolution of the chemical species over time) with the real chemical path. The integrated absolute loss is given by
    \begin{equation}\label{eq:iABS}
        \boldsymbol{L}_{\rm iABS} = \sqrt{\int_t\Big( \underline{\n}(t')dt'-\underline{\hat{\n}}(t')dt'\Big)^2},
    \end{equation}
    where the integral goes over a time interval $[1,m]$. If $m=1$, the integrated absolute loss reduces to the local absolute loss given in Eq.\ \eqref{eq:ABS}. 
\end{itemize}
For each loss, we take the norm during training (since they are vectors containing the loss per chemical species) and sum over the different samples to obtain a mean squared error (MSE) per loss type:
\begin{equation}\label{eq:MSE}
    {L}_{\rm type} = \frac{1}{N}\sum_{i=1}^N \left(\boldsymbol{L}_{{\rm type},i}\right)^2,
\end{equation}
where the index $i$ goes over the different samples in the dataset.
\\ \indent To optimise the free parameters in \mace, we follow two different schemes and compare in Sect.\ \ref{sect:results} which of the two schemes provides the proper method to predict the evolution of chemical abundances in a dynamical environment.
\begin{enumerate}
    \item \emph{Local training scheme:} training \mace\ on local information only. In this scheme, the total loss is defined as the sum of the local absolute loss (Eq.\ \ref{eq:ABS}), the local gradient loss (Eq.\ \ref{eq:GRD}), and the identity loss (Eq.\ \ref{eq:IDN}), each weighted with a factor, below given by $\lambda$, as follows: 
    \begin{equation}\label{eq:totloss_loc}
        {L}_{\rm tot}^{\rm local} = \lambda_{\rm ABS} {L}_{\rm ABS} +  {\rm GRD} {L}_{\rm GRD} + \lambda_{\rm IDN}{L}_{\rm IDN}.
    \end{equation}
    The weights $\lambda$ allow to boost certain loss types, in order to enforce certain behaviour of the model. After every timestep during the training, the free parameters of the model are updated. This method is visualised in the upper panel of Fig.\ \ref{fig:vis_loss}: during the training, this loss tries to match the coloured data points (predicted) to the black curve (real), according to Eq.\ \eqref{eq:totloss_loc}. 
    \item \emph{Integrated training scheme:} training \mace\ using the integrated absolute loss (Eq.\ \ref{eq:iABS}), hence taking into account the chemical evolution pathways. Since we still aim for the autoencoder to accurately encode and decode the data, also the identity loss (Eq.\ \ref{eq:IDN}) is included during the integrated training:
    \begin{equation}\label{eq:totloss_int}
        {L}_{\rm tot}^{\rm integrated} = {L}_{\rm iABS} + {L}_{\rm IDN},
    \end{equation}
    where the losses are not weighted. The free parameters of the model are only updated after $m$ timesteps during the training. The middle and bottom panels of Fig.\ \ref{fig:vis_loss} visualise this training scheme for $m=16$ and $m=64$, respectively. In this scheme, the predicted chemical path of length $m$ is produced at every timestep, starting from the real abundance, using the \mace\ architecture (coloured curves). These predictions appear as ``hairs'' on top of the real abundance profile (black curves). The integrated absolute loss tries to match these predicted paths (the ``hairs'') to the real chemical path (black curves), according to Eq.\ \eqref{eq:totloss_int}.
\end{enumerate}

\subsection{Hyperparameters}\label{sect:hyperparames}
The chosen architecture results in a large number of hyperparameters, i.e., parameters specific to the \mace\ architecture and training. Optimising for hyperparameters is a trial and error process, there is in principle no common way to do this. Moreover, it is possibly a degenerate problem, where changing different hyperparameters can have the same effect on the training and its results. Tools exist to do this in a systematic way. However, since this work serves a proof-of-concept, fully optimising the hyperparameters is beyond the scope of this work. 
\\ \indent Therefore, we chose to keep a number of hyperparameters fixed, namely the number of layers and nodes in the autoencoder (see Sect.\ \ref{sect:ae} and Table \ref{tab:ae}) and the learning rate. We test a limited amount of values for others, namely the dimensionality $d$ of the latent space in both schemes, the weights of the losses $\lambda$ in the local training scheme (Eq.\ \ref{eq:totloss_loc}), and the number of timesteps  $m$ in the integrated training scheme.

\subsection{Training strategy}\label{sect:trainstrat}
The Adam optimiser \citep{Kingma2017}, a stochastic gradient descent method, is used to train \mace. It is a popular choice for training neural networks, since it is computationally efficient, has little memory requirements, and is well suited for problems with large datasets and/or parameters. Moreover, it is an adaptive optimiser that adapts its stepsize in a particular way during gradient descent.
\\ \indent The different \mace\ models are trained for multiple epochs, where one epoch is defined as one full pass of the training and validation dataset. We chose to train for 100 epochs in the local training scheme (Eq.\ \ref{eq:totloss_loc}), and for 150 epochs in the integrated training scheme (Eq.\ \ref{eq:totloss_int}), since the latter is more complex and is expected to need more epochs for the model to converge. At the start of the training, the losses included in the total loss differ by orders of magnitude. Therefore, after the first 5 epochs, we rescale them to unity in order to not favour a specific loss and skew the training. Subsequently, in the local training scheme, we scale each loss according to its weight $\lambda$. 
\\ \indent In the local training scheme, we train four \mace\ architectures, each with a different latent dimensionality $d$ and loss weights $\lambda$. The hyperparameters of the local \mace\ model, named $\ell oc$, are given in Table \ref{tab:local-models}. 
The integrated training scheme introduces an extra hyperparameter, namely the number of timesteps $m$ of the chemical pathway. A large enough number of timesteps is preferred to enforce stability over a long enough period of time. However, introducing more timesteps will increases the computational cost of the training. We chose to train for three values: $m \in [8,16,32]$. Also, three different latent dimensionalities are used: $d \in [8,16,32]$. A larger latent dimensionality will result in more free parameters to train\footnote{\label{fn:freeparams}For both the local and integrated models, when latent dynamisonality $d=8$, the number of free parameters equals 284\,692, for $d=16$ the number of free parameters is 289\,508, and for $d=32$ the number of free parameters is 321\,028, giving an increase of about 36\,000 between the former and the latter.}, also increasing the computational cost of the training. Hence, by combining these values for $m$ and $d$, we strive to find a balance between training time and accuracy. The combination of the different hyperparameters results in a grid of nine integrated \mace\ models, named $\emph{int}$, given in Table \ref{tab:int-models}.
\begin{table}
    \begin{center}
        \caption{Hyperparameters for the local \mace\ models, with $d$ the dimensionality of the latent space, and $\lambda$ the weights of the losses, as defined in Eq.\ \eqref{eq:totloss_loc}. These models are trained for 100 epochs with an initial learning rate of $10^{-4}$.}
        \begin{tabular}{ c c c c c  }
            \hline \hline \\[-2ex]
        Model name & $d$ &  $\lambda_{\rm ABS}$ & $\lambda_{\rm GRD}$ & $\lambda_{\rm IDN}$ \\  \hline       
        $\ell oc$\emph{1} & 8 & 1 & 1 & 1 \\
        $\ell oc$\emph{2} & 8 & $10^4$ & $10^2$ & $10^2$ \\
        $\ell oc$\emph{3} & 16 & 1 & 1 & 1 \\
        $\ell oc$\emph{4} & 16 & $10^4$ & $10^2$ & $10^2$ \\ \hline
        \end{tabular}
        \label{tab:local-models}
    \end{center}
\end{table}
\begin{table}
    \begin{center}
        \caption{Hyperparameters for the integrated \mace\ models, with $m$ the number of timesteps and $d$ the dimensionality of the latent space. All integrated models are trained for 150 epochs with an initial learning rate of $10^{-4}$.}
        \begin{tabular}{ c c c    }
            \hline \hline \\[-2ex]
        Model name      &    $m$      &  $d$       \\  \hline       
        \emph{int1}     &     8       &   8        \\
        \emph{int2}     &     8       &   16       \\
        \emph{int3}     &     8       &   32       \\
        \emph{int4}     &     16      &   8        \\
        \emph{int5}     &     16      &   16       \\
        \emph{int6}     &     16      &   32       \\
        \emph{int7}     &     32      &   8        \\
        \emph{int8}     &     32      &   16       \\
        \emph{int9}     &     32      &   32       \\   \hline
        \end{tabular}
        \label{tab:int-models}
    \end{center}
\end{table}
\\ \indent The \mace\ models are trained on a random subset of 10\,000 \csone\ models (out of the 18\,000), hence $1.34\times 10^{6}$ \cszero\ samples, with a split of 70\%--30\% for training and validation data, respectively.

\section{Results}\label{sect:results}
In this section, we show the results of trained \mace\ models with different hyperparameters and training strategies. We distinguish between the local (Sect.\ \ref{sect:local-training}) and integrated training schemes (Sect.\ \ref{sect:integrated-training}), each defined by their respective total loss function, given in Eqs. \eqref{eq:totloss_loc} and \eqref{eq:totloss_int}. 
\\ \indent In order to test the trained \mace\ models, we apply them to a test dataset, containing 3\,000 \csone\ models, hence $3\,000\times 134\approx 4 \times 10^{5}$ \cszero\ samples, separate from the training and validation set. We chose to apply the following metric on the test dataset for comparing the results from different models:
\begin{equation}\label{eq:errorlog}
    {\rm error} = \frac{\log_{10}\n-\log_{10}\hat{\n}}{\log_{10}\n},
\end{equation}  
which is executed element-wise and subsequently summed over the different chemical species. The metric states the relative ``distance'' in log-space between the true abundance $\n$ and the predicted abundance $\hat{\n}$.

\subsection{Local training scheme}\label{sect:local-training}
\begin{figure}
    \centering
    \includegraphics[width=0.475\textwidth]{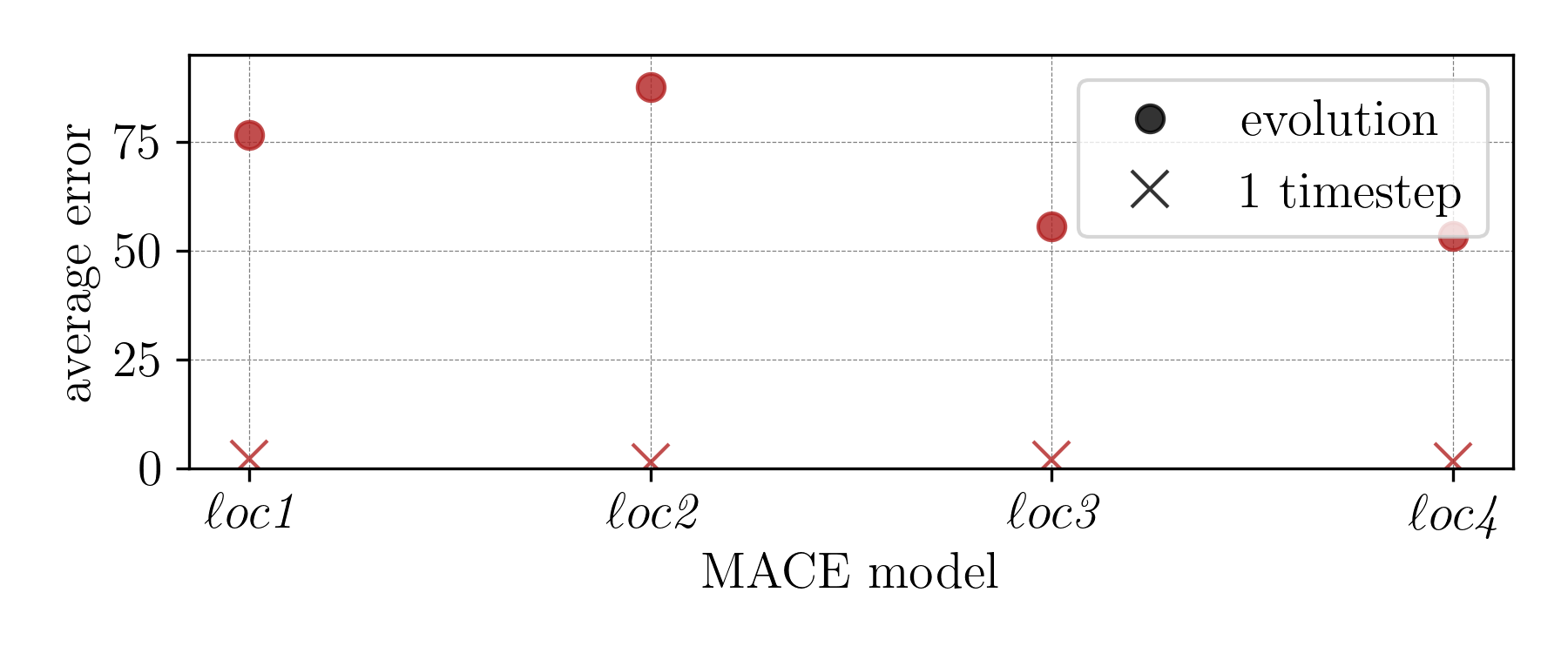}
    \caption{Average error (Eq.\ \ref{eq:errorlog}) on $4\times10^5$ test samples for the four local \mace\ models (Table \ref{tab:local-models}). The dots give the error for testing the prediction of the full chemical evolution, the crosses give the error for testing the prediction of 1 consecutive timestep only.}
    \label{fig:all_local}
\end{figure}
\begin{figure}
    \centering
    \includegraphics[width=0.475\textwidth]{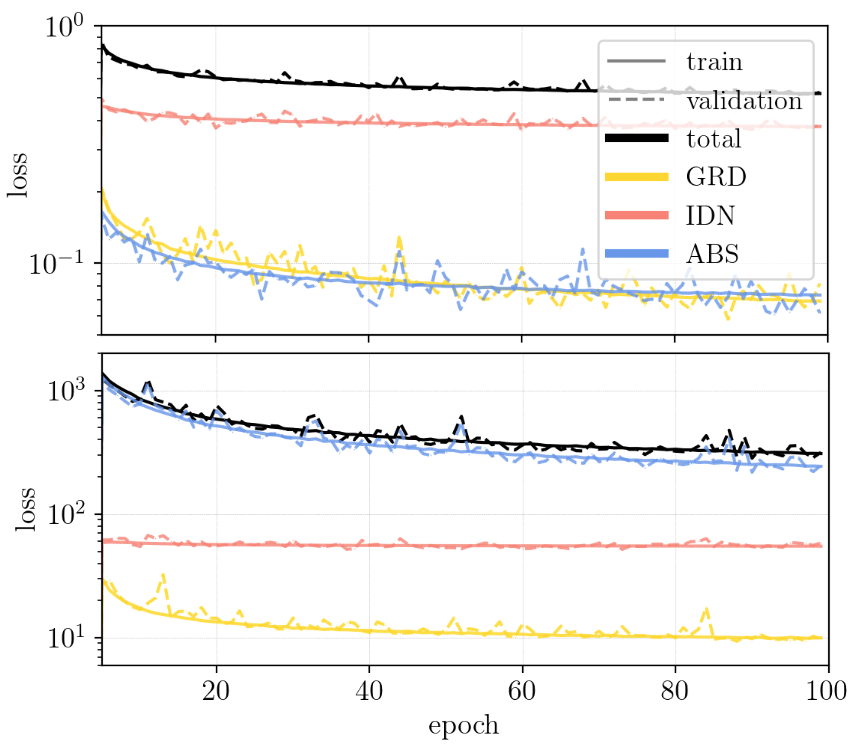}
    \caption{Losses per training epoch for model $\ell oc$\emph{3} (\underline{upper panel}) and $\ell oc$\emph{4} (\underline{bottom panel}). In colour are the individual losses (as indicated in the legend, abbreviated in Sect.\ \ref{sect:loss}). The black line gives the total loss, as defined in Eq.\ \eqref{eq:totloss_loc}. The loss on the training data is given in full lines, the loss on the validation data in dashed.}
    \label{fig:ell_loss}
\end{figure}
\begin{figure}
    \centering
    \includegraphics[width=0.5\textwidth]{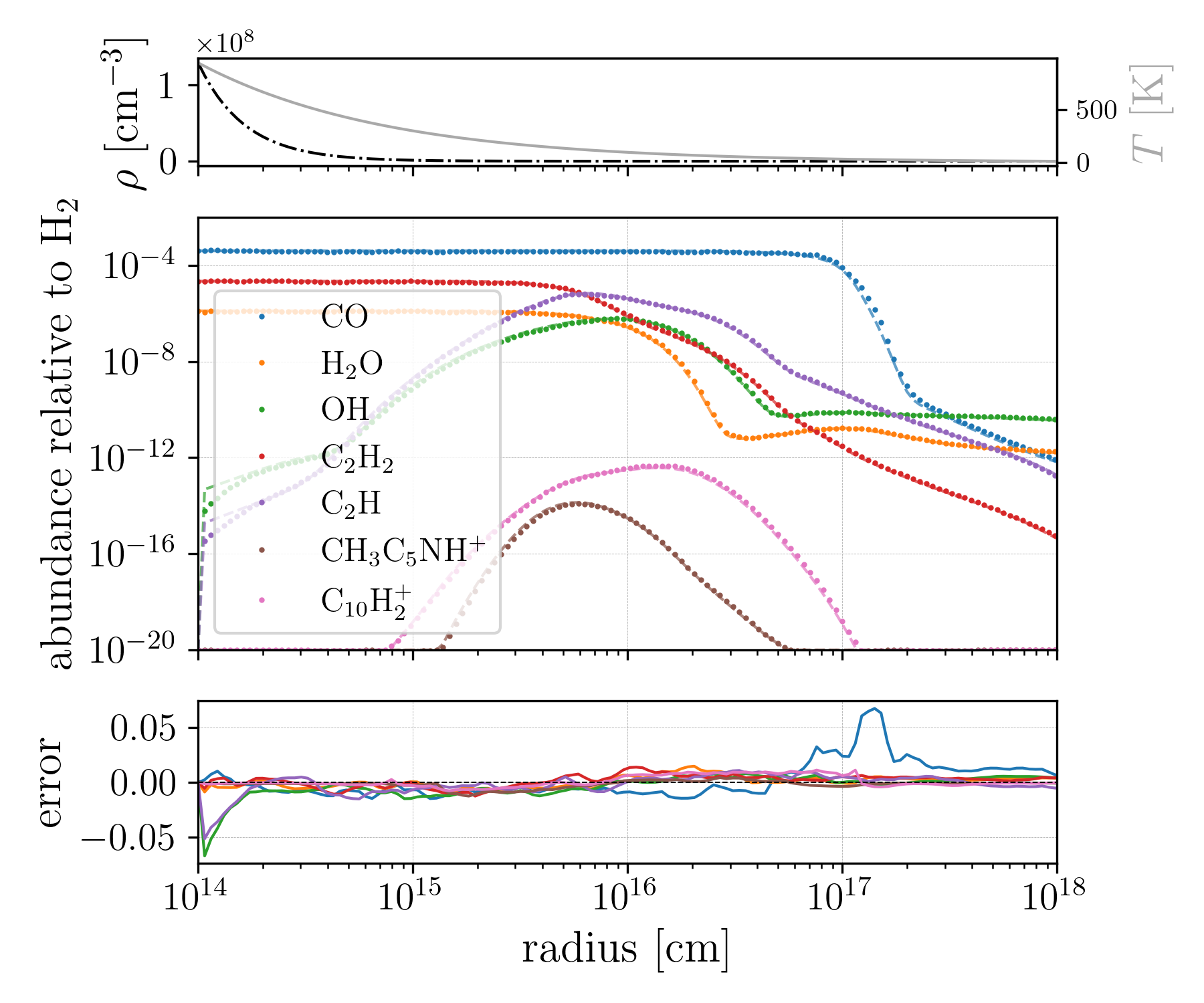}
    \caption{Timestep test of model $\ell oc$\emph{3} on a \csone\ models with the following input parameters: $\Mdot=1\times10^{-6}\,\Msolyr$, $\vexp=17.5\,\kms$, $\Tstar=2300\,$K, and $\eps=0.55$. \underline{Upper panel:} H$_2$ number density (dashed-dotted, left $y$-axis) and temperature (full grey, right $y$-axis) as a function of outflow radius.. \underline{Middle panel:} Abundances of specific species, given in legend. The dashed line gives the result of the classical model (ground truth), the dots gives the result of \mace\ only used on 1 timestep, and shown in consecutive order according to the physical parameters of the CSE. \underline{Lower panel:} Error (Eq.\ \ref{eq:errorlog}) of the \mace\ model compared to the classical model.}
    \label{fig:ell2_1step}
\end{figure}

In this section, we discuss the four \mace\ models trained according to the local scheme (Eq.\ \ref{eq:totloss_loc}), given in Table \ref{tab:local-models}. Fig.\ \ref{fig:all_local} gives the values of the error metric defined in Eq.\ \eqref{eq:errorlog}, averaged, when applying the \mace\ models to the test dataset of $4 \times 10^{5}$ \cszero\ samples for predicting the full chemical evolution path (dots), and for predicting only the abundances at the consecutive timestep (crosses). Overall for the evolution prediction, this error is found to be rather large for the four different models. This indicates that these \mace\ models do not improve significantly when certain loss types are made to dominate. When the latent dimensionality is increased, the error is slightly lower. From the average errors (Fig.\ \ref{fig:all_local}) we would prefer models $\ell oc$\emph{3} and $\ell oc$\emph{4} over the others, although not so very compelling. Fig.\ \ref{fig:ell_loss} shows the losses per epoch for model $\ell oc$\emph{3} (upper panel) and $\ell oc$\emph{4} (bottom panel). For model $\ell oc$\emph{3} it is seen that the identity loss dominates the training, which will also be the case for $\ell oc$\emph{1} (Table \ref{tab:local-models}). For model $\ell oc$\emph{4} the weight of the absolute loss compared to the gradient and identity loss is increased, making the absolute loss dominate, as will be for $\ell oc$\emph{2}. Note that in Fig.\ \ref{fig:ell_loss} the losses have stabilised by training epoch 100, indicating a minimum has been reached, i.e., the optimal solution for these architectures. Note also that because the losses are rescaled before the training by a different factor for each models, their values have no absolute meaning, thus cannot be compared between models.
\\ \indent In order to verify if a certain \mace\ model is acceptable, apart from looking only at the value of the average error on test samples, we also consider its predicted abundances. As an example, we use the \csone\ model with parameters $\Mdot=1\times10^{-6}\,\Msolyr$, $\vexp=17.5\,\kms$, $\Tstar=2300\,$K, and $\eps=0.55$ from the set of test models, resembling an average AGB outflow. To verify the evolution, we first feed the transformed input parameters of this \csone\ model, namely $\underline{\p}_0$ and $\underline{\n}_0$, together with the timestep $\Delta t_0=t_1-t_0$, to the trained \mace\ model and let it predict the next chemical state $\underline{\hat{\n}}_1$ at $t_1$. Subsequently, the \mace\ model reuses its chemical state $\underline{\hat{\n}}_1$ and we feed it the corresponding physical state $\underline{\p}_1$ from the classical model together with $\Delta t_1$. The \mace\ model then predicts the successive chemical state $\underline{\hat{\n}}_2$ at $t_2$, and so on. Hence, the successive application of \mace\ predicts the chemical evolution for a dynamical physical environment.
\begin{figure}
    \centering
    \includegraphics[width=0.5\textwidth]{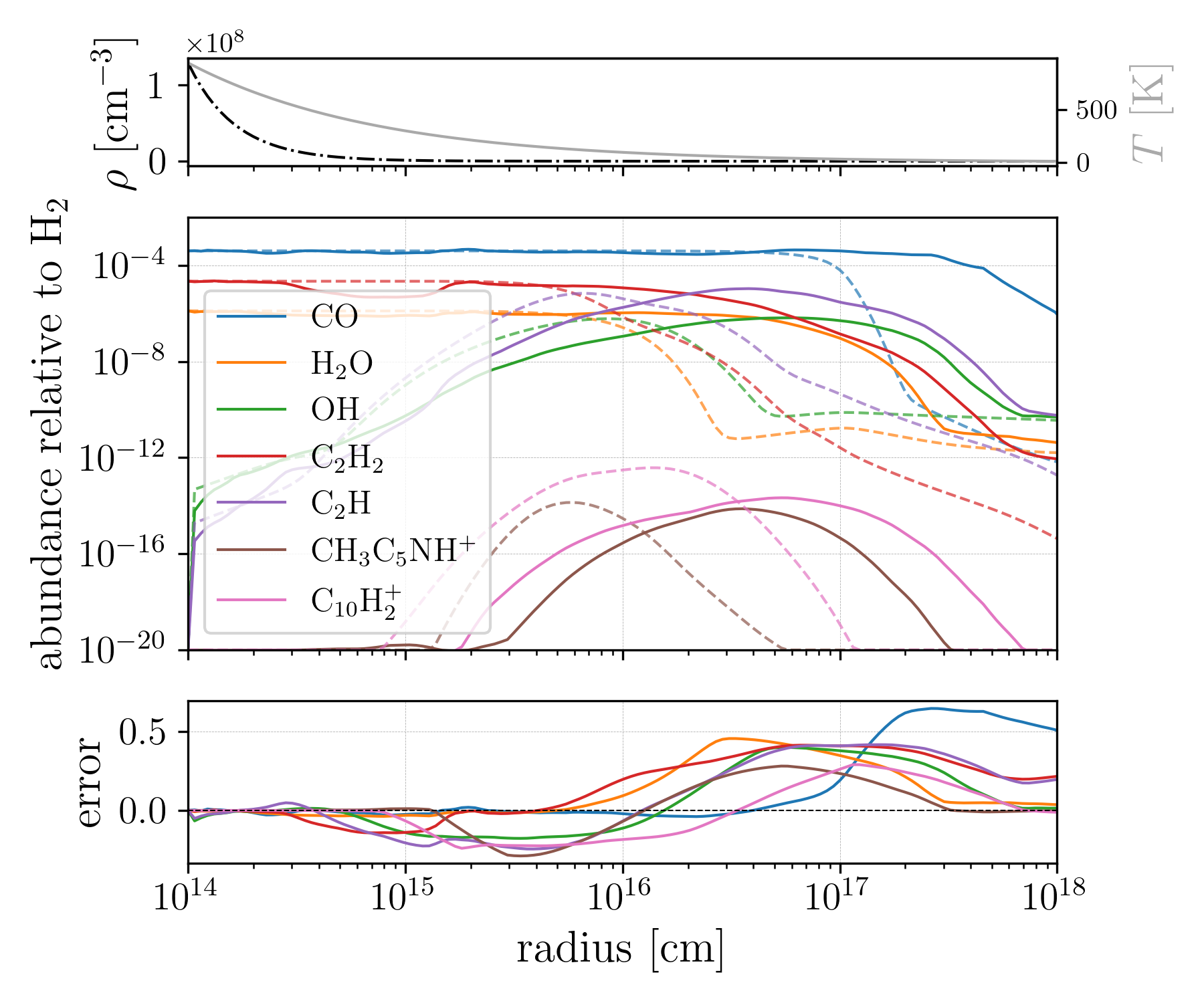}

    \caption{Evolution test of model $\ell oc$\emph{3} on a \csone\ models with the following input parameters: $\Mdot=1\times10^{-6}\,\Msolyr$, $\vexp=17.5\,\kms$, $\Tstar=2300\,$K, and $\eps=0.55$. \underline{Upper panel:} H$_2$ number density (dashed-dotted, left $y$-axis) and temperature (full grey, right $y$-axis) as a function of outflow radius. \underline{Middle panel:} Abundances of specific species, given in legend. The dashed line gives the result of the classical model (ground truth), the full line gives the result of \mace. \underline{Lower panel:} Error (Eq.\ \ref{eq:errorlog}) of the \mace\ model compared to the classical model.}
    \label{fig:ell2}
\end{figure}
\\ \indent As an example, we elaborate on the results of model $\ell oc$\emph{3}. Fig.\ \ref{fig:ell2} shows the predicted chemical evolution in full lines, and in dashed the classical analogue (taken here as ``ground truth''). The middle panel shows the abundances, relative to H$_2$, of three parent species (CO, H$_2$O, and C$_2$H$_2$), two first-generation daughter species (OH and C$_2$H), and two later-generation daughter species (CH$_3$C$_5$NH$^+$ and C$_{10}$H$_2^+$). The choice of displayed species is rather random within each category, however, the \mace\ architecture will not distinguish. The bottom panel shows the error metric, defined in Eq.\ \eqref{eq:errorlog}, per species. Although \mace\ works with time as a parameter, in these figures time is converted back to outflow radius. We see that model $\ell oc$\emph{3} is able to predict the abundances of the parent species quite well until the abundance starts to drop. This is not very surprising, because up until that point, the model should not change the initial value much, since ``doing nothing'' is easy. However, further down the evolutionary path, we see that the \mace\ model diverges strongly from the ground truth. For the daughter species, this \mace\ model is not able to reproduce their chemical path. Also the other local models are not able to reproduce of the chemical pathway, figures are given in Appendix \ref{sect:loc_app}. 
\\ \indent Although the local \mace\ models fail to reproduce the chemical evolution, they do accurately reproduce 1 consecutive chemical state (1 timestep only). We test this by feeding the \mace\ model the physical state $\underline{\p}_i$ and \emph{true} chemical state $\underline{\n}_i$ of the corresponding \csone\ model at every timestep $t_i$, and let it predict the next chemical state $\underline{\hat{\n}}_{i+1}$ at $t_{i+1}$. An example of such a test is shown in Fig.\ \ref{fig:ell2_1step} for model $\ell oc$\emph{3}, where the abundance predictions (dots) are shown in consecutive order according to the radius of the CSE. Within the chosen metric (Eq.\ \ref{eq:errorlog}), the error here is an order of magnitude lower than when predicting the full chemical evolution, as is the case for the average error over the test dataset for the different local models (see crosses in Fig.\ \ref{fig:all_local}). %This is already an impressive result, since nowhere in the literature could they accurately predict the next chemical step for such a large chemical network.
This is an impressive results, since, to the best of our knowledge, it is the first time that the next chemical step for such a large chemical network can be accurately predicted in a machine learning approach. However, this leaves us with the challenge to robustly predict the chemical evolution in a dynamical environment, which we address in the next section.

\subsection{Integrated training scheme}\label{sect:integrated-training}
\begin{figure}
    \centering
    \includegraphics[width=0.46\textwidth]{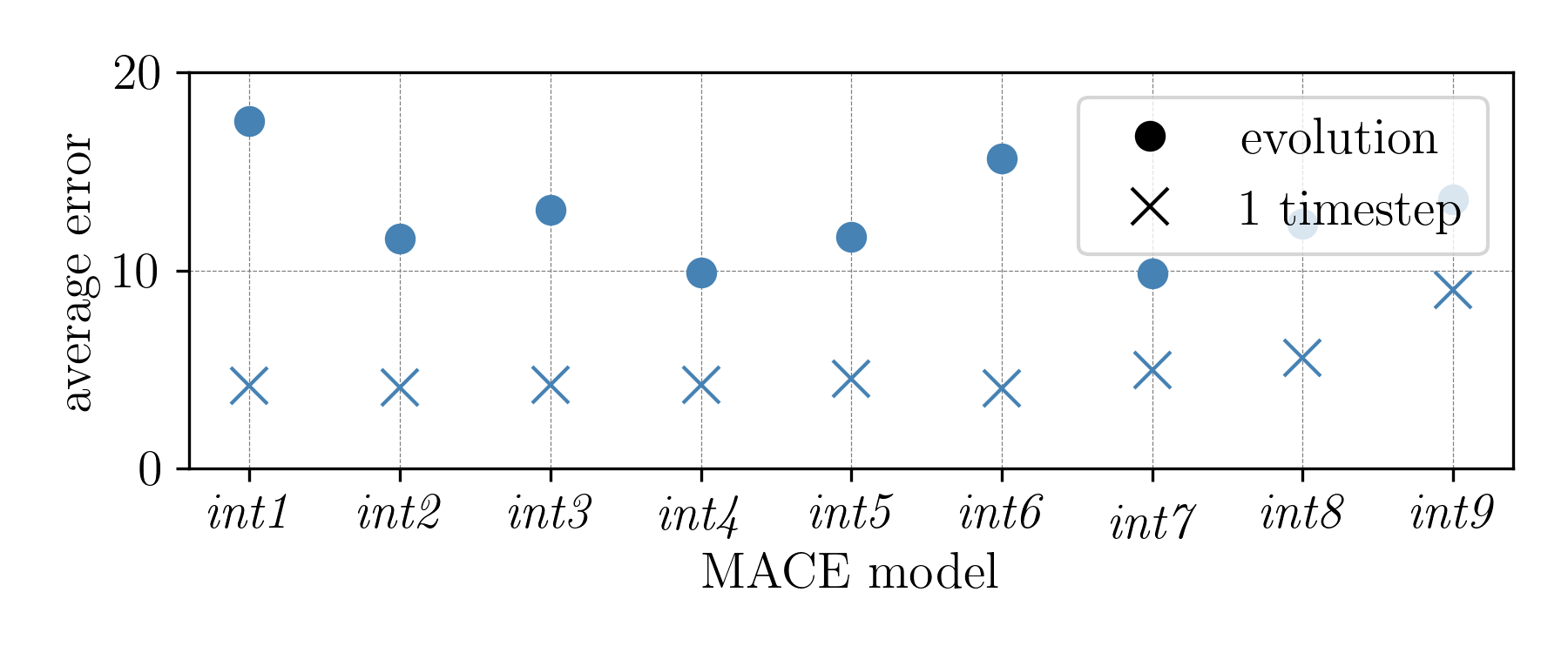}
    \caption{Average error (Eq.\ \ref{eq:errorlog}) on $4\times10^{5}$ test samples for the integrated \mace\ models (Table \ref{tab:int-models}). The dots give the error for testing the prediction of the full chemical evolution, the crossing give the error for testing the prediction of 1 consecutive timestep only.}
    \label{fig:all_integrated}
\end{figure}
\begin{figure}
    \centering
    % \vspace{-0.25cm}
    % \includegraphics[width=0.475\textwidth]{/STER/silkem/MACE/models/CSE_0D/20240207_134859_66656_2/plots/loss.png}
    % \includegraphics[width=0.45\textwidth]{20240207_134859_66656_2-loss.png}
    % \vspace{-0.25cm}
    % \includegraphics[width=0.475\textwidth]{/STER/silkem/MACE/models/CSE_0D/20240207_134859_66656_4/plots/loss.png}
    % \includegraphics[width=0.475\textwidth]{figs/66656_4_loss.png}
    % \vspace{0.25cm}
    % \includegraphics[width=0.475\textwidth]{/STER/silkem/MACE/models/CSE_0D/20240208_135604_66879_7/plots/loss.png}
    % \includegraphics[width=0.45\textwidth]{20240208_135604_66879_7-loss.png}
    \includegraphics[width=0.475\textwidth]{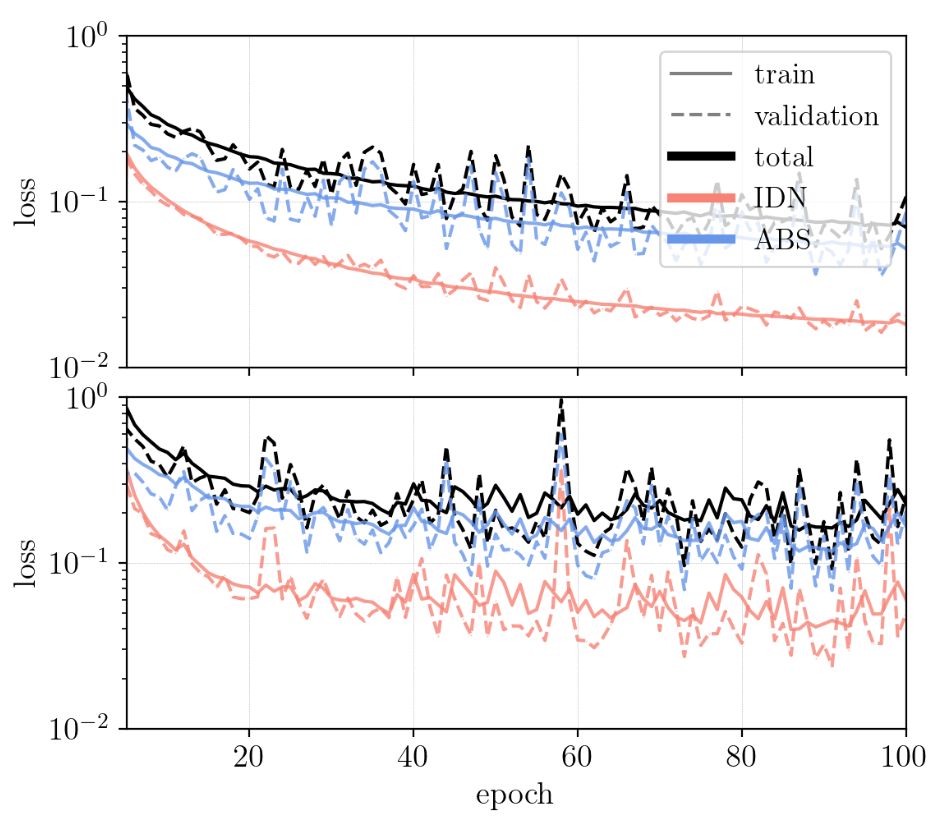}
    % \vspace{-0.25cm}
    \caption{Losses per training epoch for model \emph{int2} (\underline{upper panel}) and \emph{int7} (\underline{bottom panel}). In colour are the individual losses (as indicated in the legend, abbreviated in Sect.\ \ref{sect:loss}). The black line gives the total loss, as defined in Eq.\ \eqref{eq:totloss_loc}. The loss on the training data is given in full lines, the loss on the validation data in dashed.}
    \label{fig:int_loss}
\end{figure}

In this section, we discuss the \mace\ models trained according to the integrated scheme (Eq.\ \ref{eq:totloss_int}), given in Table \ref{tab:int-models}. Fig.\ \ref{fig:all_integrated} shows the error metric (Eq.\ \ref{eq:errorlog}), averaged over the same test dataset of $4\times10^5$ samples, after applying the \mace\ models on it. Within this metric, the average errors are about a factor 4 smaller than the average errors on the local models (Fig.\ \ref{fig:all_local}). This is a significant improvement. Again, the errors of the different models lie close to each other, from which we cannot clearly prefer one over the other. Fig.\ \ref{fig:int_loss} shows examples of the loss per epoch for models \emph{int2} (upper panel) and \emph{int7} (bottom panel). We see that for model \emph{int7} the training loss (full lines) does not decrease with every epoch, giving it a spiky look. This model is more complex, containing many more free parameters than, e.g., \emph{int2} (see footnote \ref{fn:freeparams}). Therefore, it is harder to find a local minimum, especially when the learning rate is large. We suspect that training this model with a lower learning rate will smoothen the train loss curve. Though, the general trend of the training loss is a decrease, indicating that the model does converge. 
\begin{figure}[h]
    \centering
    \includegraphics[width=0.5\textwidth]{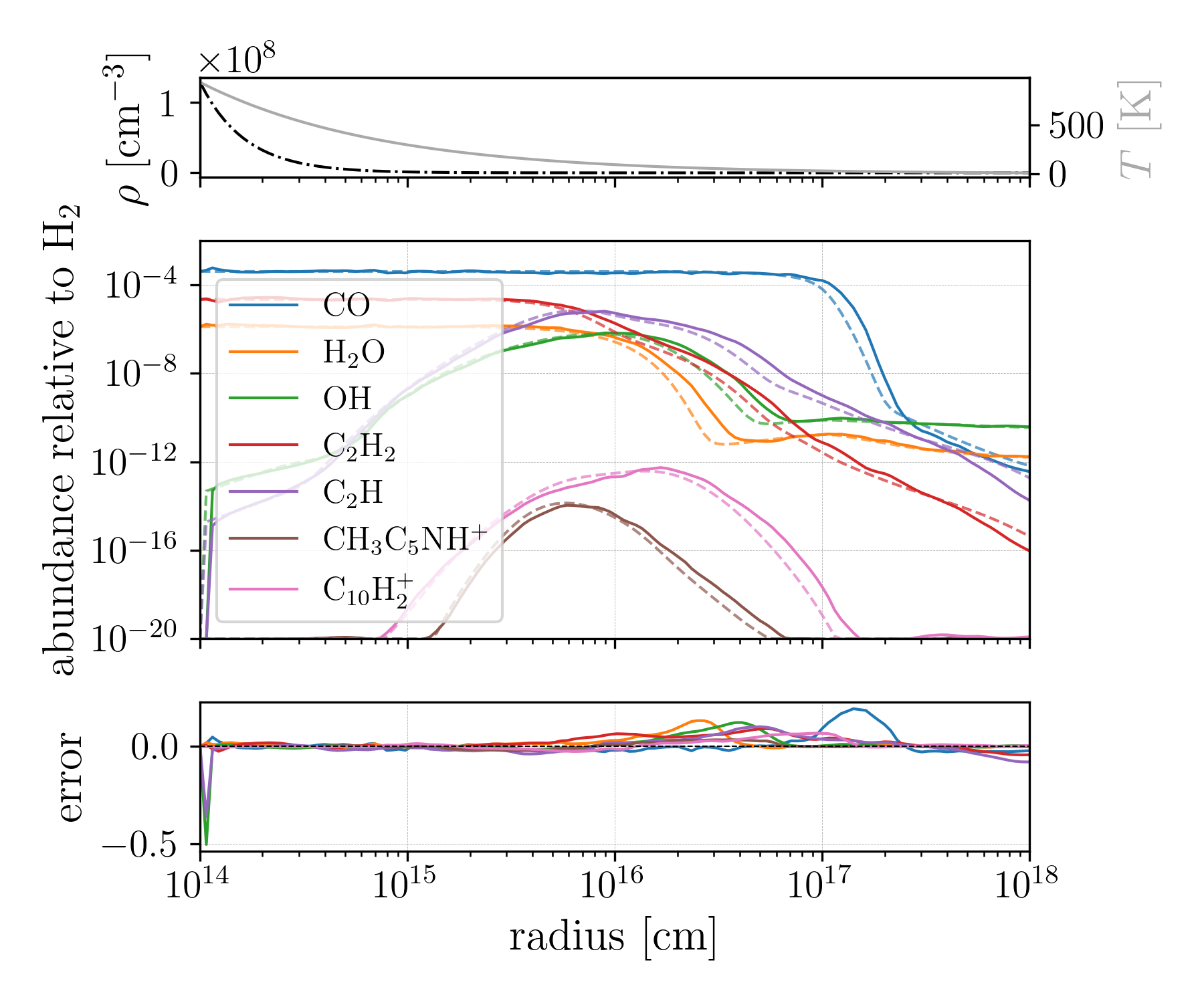}
    \caption{Evolution test of model \emph{int4} on a \csone\ models with the following input parameters: $\Mdot=1\times10^{-6}\,\Msolyr$, $\vexp=17.5\,\kms$, $\Tstar=2300\,$K, and $\eps=0.55$. \underline{Upper panel:} H$_2$ number density (dashed-dotted, left $y$-axis) and temperature (full grey, right $y$-axis) as a function of outflow radius. \underline{Middle panel:} Abundances of specific species, given in legend. The dashed line gives the result of the classical model (ground truth), the full line gives the result of \mace. \underline{Lower panel:} Error (Eq.\ \ref{eq:errorlog}) of the \mace\ model compared to the classical model.}
    \label{fig:int4_abs}
\end{figure}
\\ \indent Fig.\ \ref{fig:int4_abs} shows the predicted chemical evolution by model \emph{int4} on the same example \csone\ test model. Compared to the results of the local models (Sect.\ \ref{sect:local-training}), these findings are a great improvement. The predicted chemical evolution almost matches the real evolution exactly; the \mace\ model is able to reproduce the chemical pathway of the parent and daughter species. The other integrated models give similar, good results. Figures can be found in Appendix \ref{sect:integrated_app}. 
\\ \indent Generally, the integrated \mace\ models perform better on a high density outflow. For lower density models, a systematic offset is noticed in the predicted abundances by \mace\, (see e.g., left panel of Fig.\ \ref{fig:int4_low} in Appendix \ref{sect:integrated_app}). This is due to the effect of the CO self-shielding, which is depending on the velocity of the outflow and can affect the abundances significantly \citep{Maes2023}. However, since a \mace\ model receives the density as input (Eq.\ \ref{eq:physpar}) and not the mass-loss rate and expansion velocity separately (contrary to the classical model), this chemical subtlety is not grasped by \mace. Though, this can easily be included in an improved version of \mace, as demonstrated by this proof-of-concept.

\section{Discussion}\label{sect:discussion}
In this section, we elaborate on the integrated \mace\ models, since the integrated training scheme is found to be succesful, contrary to the local scheme. We discuss the accuracy of their results on test data, as a function of their training time. Moreover, we determine the speed-up of using \mace\ instead of the classical CSE model from Sect.\ \ref{sect:testcase}. Finally, we discuss possible improvements of the \mace\ architecture and training.

\subsection{Accuracy \& training time}\label{sect:accuracy}
\begin{figure}
    \centering
    \includegraphics[width=0.475\textwidth]{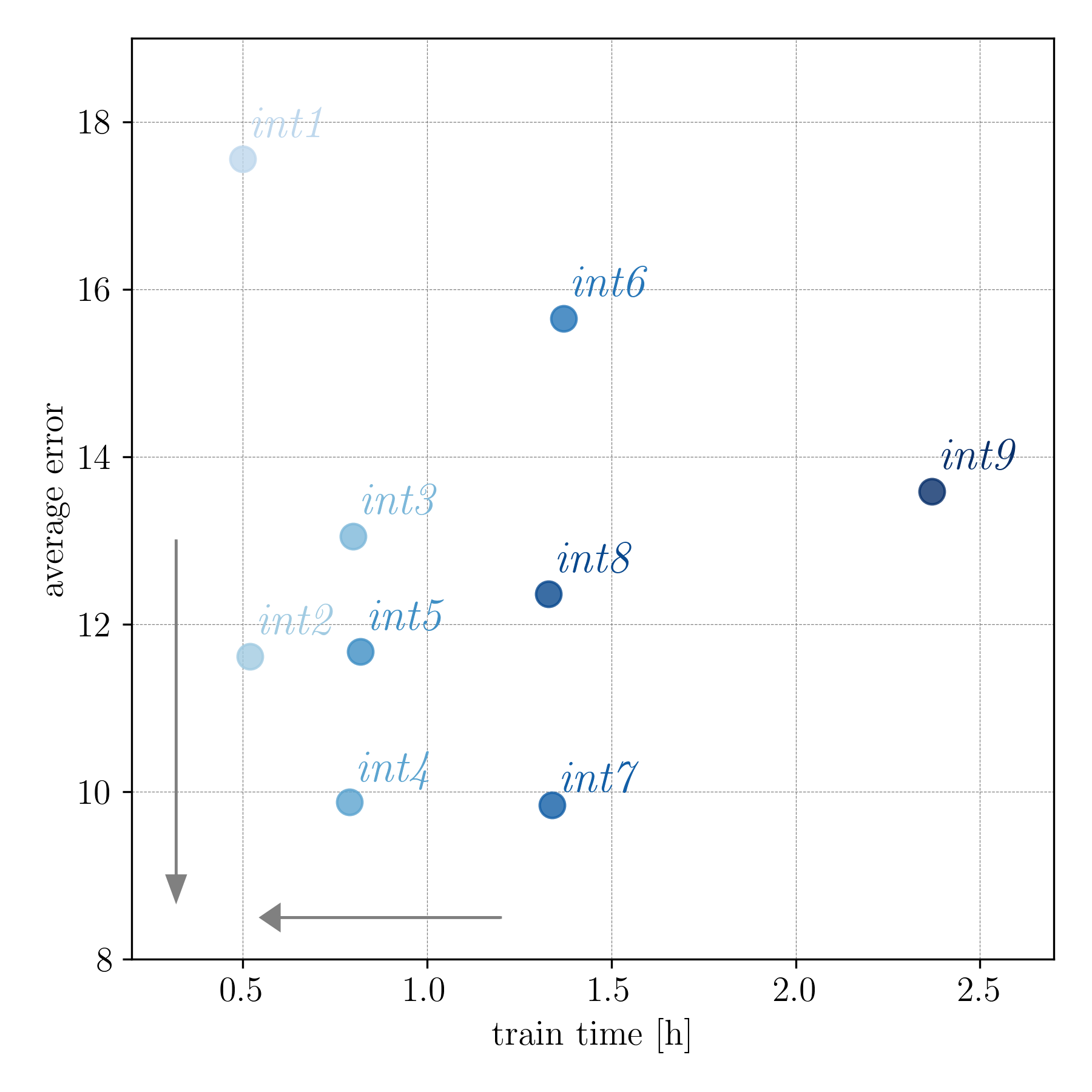}
    \caption{Performance landscape of integrated \mace\ models: accuracy (given by mean error, Eq.\ \ref{eq:errorlog}) as a function of training time needed to complete one epoch. The arrows indicate the direction of improvement in this landscape of the preferred models. The darker the colour, the more complex the model.}
    \label{fig:acc_traintime}
\end{figure}

Training a more complex model generally takes more time. As such, it is only beneficial to train a more complex model when it results in a significant improvement in accurately. Fig.\ \ref{fig:acc_traintime} shows the performance landscape of our models; the mean error (Eq.\ \eqref{eq:errorlog}, used here as a measure of accuracy) of the integrated \mace\ models is given as a function of the time it took to train one epoch\footnote{Training times of the different models should be compared only relatively, since, depending on which and how many CPUs the training is performed, the absolute values will differ from what is given here. The models in this work are all trained on the same machine, using the same amount of CPUs}. The arrows indicate the direction of improvement within this landscape. The more complex models (i.e., models containing more free parameters and/or using more steps $m$ in their training scheme) are indicated with a darker shade of blue. We find that they do not necessarily perform better than the simpler models (e.g., model \emph{int9} versus model \emph{int3}). More specifically, this indicates that using more timesteps $m$ in the integrated training scheme is not necessary. Moreover, we find that the models with a smaller latent dimensionality $d$ (e.g., \emph{int2}, \emph{int4}, and \emph{int7}) give a better performance, as they have a better accuracy for a shorter training time.
\\ \indent Besides, there is a caveat regarding the performance of chemistry emulators. They are trained on data coming from classical chemistry models, in this case believed to be the ``ground truth''. However, also the classical models give a certain error compared to reality, since they rely on reaction rates gathered from experiments or estimated from theory. Assessing the error on classical models via sensitivity analysis is an active field of research (e.g., Van de Sande, \emph{in prep}) and ballpark values are yet unknown. Hence, there is no use in trying to reduce the mean error on \mace\ to zero. Accordingly, we claim \emph{int2} to be our best performing model in this proof-of-concept work, since it is located in the performance landscape amongst the lowest mean errors and, especially, has the shortest training time.

\subsection{Speed-up}\label{sect:speedup}
The ultimate goal of building a chemical emulator is to speed up the chemistry computations. In this section, we elaborate on the speed-up that \mace\ offers compared to the classical CSE model. We  time how long it takes the solvers (\mace\ versus the classical model, the latter using \textsc{dvode} from \textsc{odepack} \citealp{VODE,ODEPACK}) to calculate the evolution of the abundances of a certain \csone\ model, and as such exclude any overhead time. We distinguish between a \csone\ model with a low and high density, as we know the computation of the \textsc{dvode} solver will differ in both cases. The results are given in Table \ref{tab:speed-up}.
\begin{table}
    \begin{center}
        \caption{Computation time to produce a \csone\ chemistry model with \mace\ and the classical \textsc{dvode} solver. The computation time is given in seconds.}
        \begin{tabular}{ r c c c c }
            \hline \hline \\[-2ex]
        computation time [s]& \mace\ & classical & & speed-up  \\  \hline       
        low density & 0.6 & 14.6 && $\times 24$ \\
        high density & 0.6  & 16.8 && $\times 28$ \\  \hline
        \end{tabular}
        \label{tab:speed-up}
    \end{center}
\end{table}
\\ \indent With \mace, we find a speed-up of a factor 24 for low-density outflows and a factor 28 for high-density outflows, since the \mace\ solver is independent of the density in the outflow. Although a speed-up of one order of magnitude for chemistry calculations might not be sufficient to make a 3D hydro-chemistry feasible, the power of \mace\ lies within the optimised use of matrix operations in its implementation with PyTorch. %As such, the chemical evolution of multiple particles in a hydrodynamical simulation can be calculated during one pass of the emulator only. This is not the case with a classical chemistry model, since then the computation time will increase linearly with the number of particles. 
As such, we can use the PyTorch framework to efficiently parallelise the computations for multiple particles. Furthermore, we can leverage hardware accelerators, such as graphics processing units (GPUs) or tensor processing units (TPUs), specifically optimised for this kind of computations. Therefore, the speed-up reported here provides merely a lower bound. The true practical speed up will depend on the particular application.

\subsection{Future prospects}\label{sect:future}
\mace\ serves as a proof-of-concept for emulating chemical evolution pathways in a dynamical environment. {As such, its architecture can be trained on chemical models of other astrophysical environments as well, for instance protoplanetary disks, diffuse or molecular ISM, dark clouds, etc. The set of physical parameters used in this research (Eq.\ \ref{eq:physpar}), most probably, will not be ideal to use in other environments. Therefore, the physical parameters should be adapted to the specific environment, a flexibility which \mace\ is designed for (Eq.\ \ref{eq:mace}).} 
\\ \indent
Moreover, the current implementation of \mace\ can be further developed before it is applied in a hydrodynamical simulation. In this section, we discuss some of the possible improvements that can be made to the architecture and training in order to get a better performance, and elaborate on the application of \mace\ in hydrodynamical simulations.
 
\subsubsection{Possible improvements}\label{sect:improvements}
The architecture of \mace\ can be improved in different ways, from a physical/chemical point-of-view, as well as from an implementation/machine learning point-of-view. %The latter approach is two-fold. We here discuss them separately. 
\\ \indent Improving on the physics and chemistry in \mace\ can be done by including other types of (physically informed) losses in the training. For example, for an environment undergoing chemical evolution, the total mass of the environment is conserved (in the absence of nucleosynthesis). \cite{Grassi2022} and \cite{Sulzer2023} have included mass conservation as an extra loss term in there architecture. This requirement can even be made stronger, by stating that the total number of atoms of each element is conserved, thus adding element conservation. Including this can be done by adding an extra loss term to the total loss (Eq.\ \ref{eq:totloss_loc}) or by constructing the latent ODE (Eq.\ \ref{eq:latentODE}) in such a way that it is built in. In Appendix \ref{sect:elementcons}, we elaborate on this prospect. 
\\ \indent We can also improve \mace\ from a machine learning point-of-view. First of all, keeping the current architecture, the hyperparameters can be further optimised, for example by training with another learning rate and for more epochs. Secondly, the architecture can be developed further by adding more layers and/or nodes per layer to the encoder and decoder, allowing for more complexity. However, this will also add time to the training procedure. Therefore, it would be advantageous to leverage GPU acceleration, instead of CPU only. Additionally, the autoencoder and latent ODE can be trained separately, contrary to what is done in this work. This approach can be very beneficial; (i) it would allow for analysing and finetuning better the latent space and as such optimising its dynamics, and (ii) multiple timesteps can be performed in latent space without the need to decode and encode every step. This will not only save some computation time, but rather allow for economic memory use, since only a greatly reduced amount of features need to be stored. Expanding on this approach, multiple, various decoders can be developed to match the desired purpose. For example, if the purpose is to only model a set of parent species, a decoder can be trained to only return that specific set, again saving computation time. We note that, in order to separately train the two parts, the physical parameters (Eq.\ \ref{eq:physpar}) should be incorporated in the latent space and not in the encoder, altering slightly the flow of \mace\ given in Eq.\ \eqref{eq:mace}. This can be done, for example, by constructing the latent ODE coefficient tensors (Eq.\ \ref{eq:latentODE}) from neural networks taking the physical parameters as input. 
\\ \indent Moreover, improvements in the implementation can benefit and increase the speed-up \mace\ provides over classical models. Currently, \emph{torchode} \citep{lienen2022torchode} is used as ODE solver, since we need its gradient tracking to train the latent ODE system (Eq.\ \ref{eq:latentODE}). However, the solver by itself is relatively slow and currently dominates the computation time of \mace\ over the autoencoder by a factor of 10. Hence, once an optimal \mace\ model is acquired, \emph{torchode} can be substituted for an alternative ODE solver that is faster for this dynamics. Furthermore, the tolerances of the ODE solver can be optimised regarding the error between the results of the emulator and the classical model, potentially increasing even more the speed-up. 

\newpage
\subsubsection{Implementation in hydrodynamical simulations}
Now that we have established that \mace\ performs properly and is able to reproduce a 1D chemical pathway in dynamical environment, the next step is to couple the emulator to a 3D hydrodynamics simulation. Because the chemical models, on which \mace\ is currently trained, cannot deal with the advection of chemical abundances, \mace\ should be coupled to
a Lagrangian hydrodynamics model, which allows us to solve the chemistry in a co-moving reference frame, for example a smoothed particle hydrodynamics (SPH) framework \citep{Lucy1977,GandM1977}. {Moreover, in order to correctly implement the radiation-induced chemical reactions in a 3D hydrodynamical model, such as photodissociation, the radiation-related parameters, $\xi$ and $A_{\rm V}$ (Eq.\ \ref{eq:physpar}), should be calculated as well. This can either be achieved in a very approximate way, or more elaborate by the use of a ray-tracing algorithm.} %This is crucial for the study of the interplay between chemistry and hydrodynamics, and to model the physical processes in the outflow more correctly.
\\ \indent As future work, we aim to couple \mace\ to the SPH code \Phantom\ \citep{PriceFederarth2010,Price_Phantom2018} in the framework of AGB outflows \citep{Siess2021,Esseldeurs2023}. {The radiation-related parameters, $\xi$ and $A_{\rm V}$, will be calculated by using the ray tracer of the 3D radiative transfer solver \textsc{Magritte}\footnote{\textsc{Magritte} is open source and can be found online: \url{https://github.com/Magritte-code/Magritte}, \url{https://magritte.readthedocs.io/en/stable/}.} \citep{DeCeuster2020I_Magritte,DeCeuster2020II_Magritte, Magritte_JOSS2022}, similar to how it was used by \cite{Esseldeurs2023}.} The coupling will allow to generate 3D hydro-chemical models of AGB outflow perturbed by a companion \citep{Maes2022}. This would allow us to step away from a 1D-approach \citep{VdSMillar2022} and better study the impact of the companion on the chemistry in the outflow. As such, chemical signatures of hidden companions can be identified, which is crucial for the interpretation of observations of AGB outflows. Besides, the \mace\ framework provides us with the prospect of implementing chemical cooling and heating processes in the hydrodynamics simulation, and inform molecular line cooling, in order to model and study the interplay between chemistry and hydrodynamics, and more correctly model the physical processes in the outflow.

\section{Conclusion}\label{sect:conclusion}
This work presents \mace, a \emph{Machine learning Approach to Chemistry Emulation}, as a proof-of-concept for emulating chemistry in a dynamical environment. Inspired by literature findings (e.g., \citealp{Holdship2021,Grassi2022,Sulzer2023}), we have constructed an architecture (Eq.\ \ref{eq:mace}) where an autoencoder compresses the chemical network to a latent space (Eqs.\ \ref{eq:encoder} and \ref{eq:decoder}). Subsequently, in this mathematical latent space, the chemical evolution is emulated by solving the latent coupled ordinary differential equation (Eq.\ \ref{eq:latentODE}). The physical parameters of the environment (Eq.\ \ref{eq:physpar}) are included in the encoder. 
\\ \indent For the first time, it is possible to accurately predict chemical abundances for a large chemical network of 468 species and 6180 reactions. Moreover, we find that using an integrated training scheme (Eq.\ \ref{eq:totloss_int}) allows to reproduce a full chemical pathway in a \emph{dynamical} environment, something that has not been done before. As an example, we apply it to the dynamical environment of AGB star's circumstellar envelopes, with the objective to couple this chemistry emulator to existing 3D hydrodynamical models. In order for this to be feasible, \mace\ should have a fast performance.
\\ \indent We find \mace\ to have a speed-up of one order of magnitude compared to its classical analogue, complementary to the optimised use of matrix operations in its implementation with PyTorch. The latter makes that only one pass of the emulator is needed for computations of the chemistry of all particles in a mesh-free hydrodynamics simulation, contrary to the a classical chemistry model where computation will grow linearly. 
\\ \indent The current implementation of \mace\ offers opportunities for further development, such as including element conservation and refining its architecture, most likely enhancing its performance. {Moreover, \mace\ is designed to be flexible, so that it can be applied to other astrophysical environments as well, by retraining its architecture on the appropriate models.}

%% Also note that the akcnowlodgment environment does not support long amounts of text. If you have a lot of people and institutions to acknowledge, do not use this command. Instead, create a new 
\section*{Acknowledgements}
{ S.M.\ and L.D.\ acknowledge support from the Research Foundation Flanders (FWO) grant G099720N. F.D.C.\ is a Postdoctoral Research Fellow of the Research Foundation - Flanders (FWO), grant number 1253223N, and was previously supported for this research by a Postdoctoral Mandate (PDM) from KU Leuven, grant number PDMT2/21/066. M.V.d.S.\ acknowledges support from the  European Union's Horizon 2020 research and innovation programme under the Marie Sk\l odowska-Curie grant agreement No 882991 and the Oort Fellowship at Leiden Observatory. L.D.\ also acknowledges support from KU Leuven C1 MAESTRO grant C16/17/007, KU Leuven C1 BRAVE grant C16/23/009, and KU Leuven Methusalem grant METH24/012.}


%\vspace{5mm}
%\facilities{HST(STIS), Swift(XRT and UVOT), AAVSO, CTIO:1.3m,CTIO:1.5m,CXO}


\software{UMIST \csone\ model (\url{https://github.com/MarieVdS/rate22_cse_code}, \citealp{McElroy2013, Millar2024}), PyTorch \citep{pytorch},  \emph{torchode} \citep{lienen2022torchode}          }



% \newpage
\appendix
\section{Physical parameter space}\label{sect:physpar_app}
\begin{figure}[h]
    \centering
    \includegraphics[width=1\textwidth]{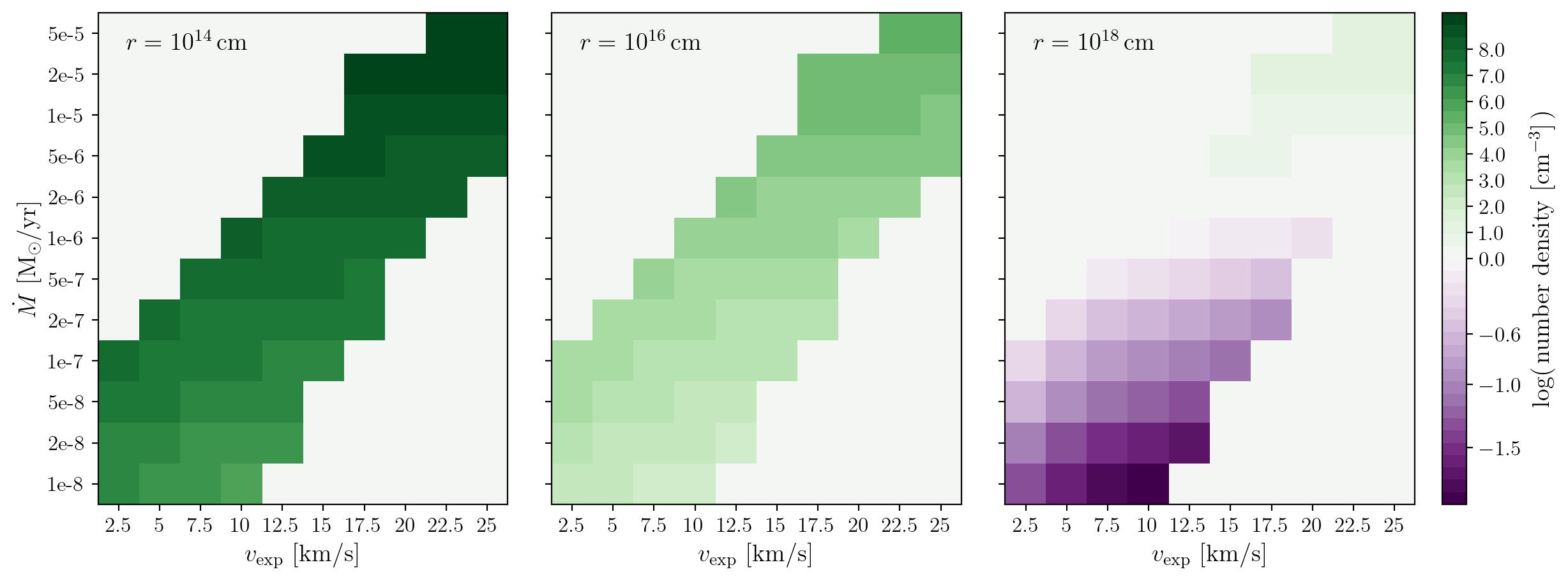}
    \caption{{Visualisation of the number density space ($\rho_N=\frac{\rho}{\mu m_H}$, with $\rho$ given in Eq.\ \eqref{eq:density}, $\mu$ the mean molecular mass per H$_2$ molecule and $m_H$ the atomic mass unit) of the training data, via the combinations of expansion velocity $\vexp$ and mass-loss rate $\Mdot$, given at different radii. (Adapted from \citealp{Maes2023})}}
    \label{fig:grid_dens}
\end{figure}

\begin{figure}[h]
    \centering
    \includegraphics[width=0.6\textwidth]{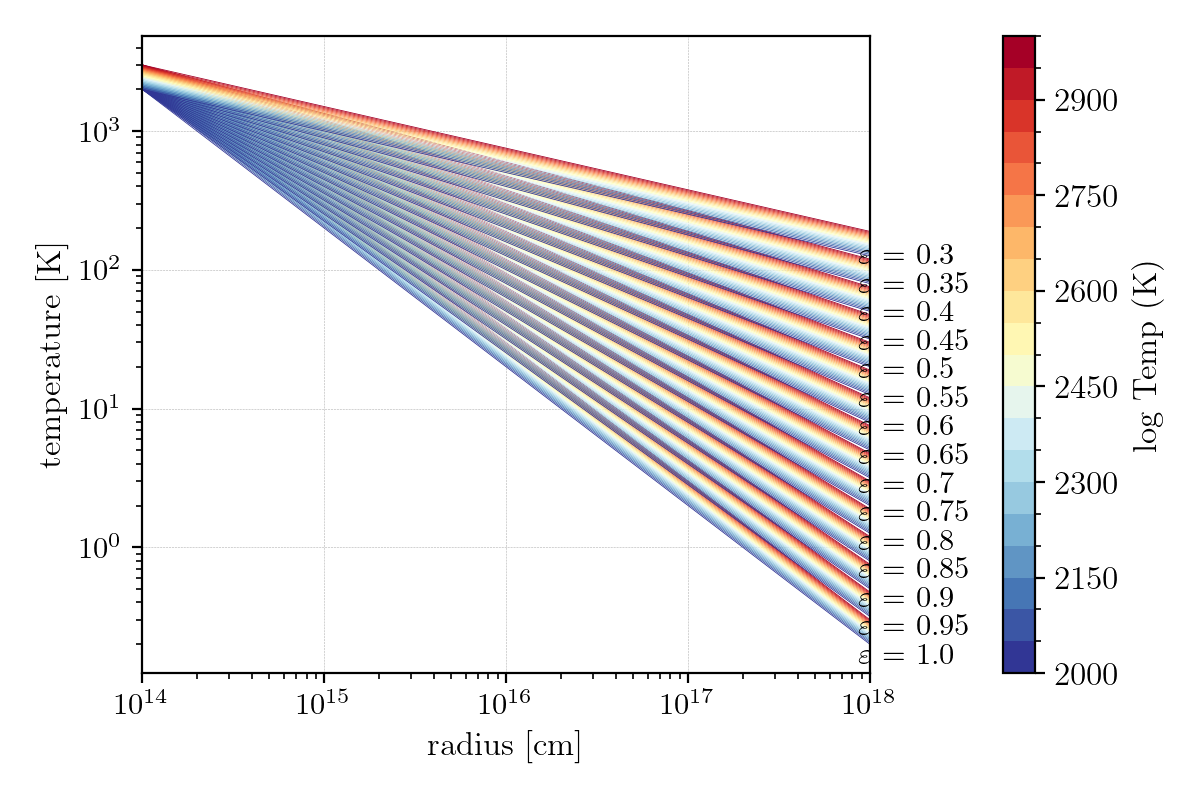}
    \caption{{Visualisation of the different temperature profiles (Eq.\ \ref{eq:temp}) in the training data, where the stellar temperature $\Tstar$ is indicated by the colourbar. The different values of $\eps$ result in different groups of temperature profiles, indicated at the right-hand side of the panel. (Adapted from \citealp{Maes2023})}}
    \label{fig:grid_temp}
\end{figure}

\begin{figure}[h]
    \centering
    \includegraphics[width=0.7\textwidth]{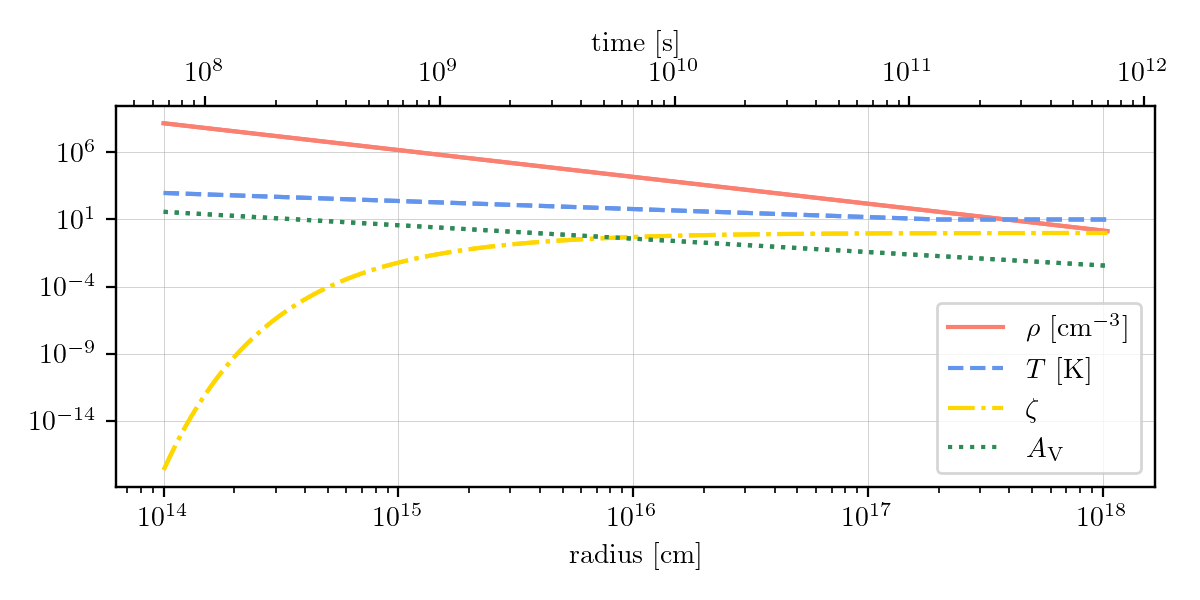}
    \caption{The physical parameters $\p=(\rho, T, \xi, \AV)$ as a function of time (upper $x$-axis) and radius (bottom $x$-axis), given for an example \csone\ models with input parameters $\Mdot=1\times10^{-6}\,\Msolyr$, $\vexp=15\,\kms$, $\Tstar=2500\,$K, and $\eps=0.6$. The $y$-axis states the value of the specific parameter given in by the legend.}
    \label{fig:physpar_expl}
\end{figure}

This section shows the physical parameter space of the classical \csone\ models. Fig.\ \ref{fig:grid_dens} shows the number density ranges ($\rho_N=\frac{\rho}{\mu m_H}$, with $\rho$ given in Eq.\ \eqref{eq:density}, $\mu$ the mean molecular mass per H$_2$ molecule and $m_H$ the atomic mass unit) for the grid of models set by mass-loss rate $\Mdot$ and expansion velocity $\vexp$, at three different locations in the outflow ($r=10^{14}\,\cm$, $r=10^{16}\,\cm$, and $r=10^{18}\,\cm$). Fig.\ \ref{fig:grid_temp} shows the temperature ranges, set by $\Tstar$ and $\eps$ in Eq.\ \eqref{eq:temp}. Fig.\ \ref{fig:physpar_expl} gives an example of the physical parameters (Eq.\ \ref{eq:physpar}) of an \csone\ models as a function of radius and time, which forms the basis of the dynamical physical environment in which the chemical reactions take place.

\section{Testing local models}\label{sect:loc_app}
\begin{figure}[h]
    \centering
    \includegraphics[width=0.475\textwidth]{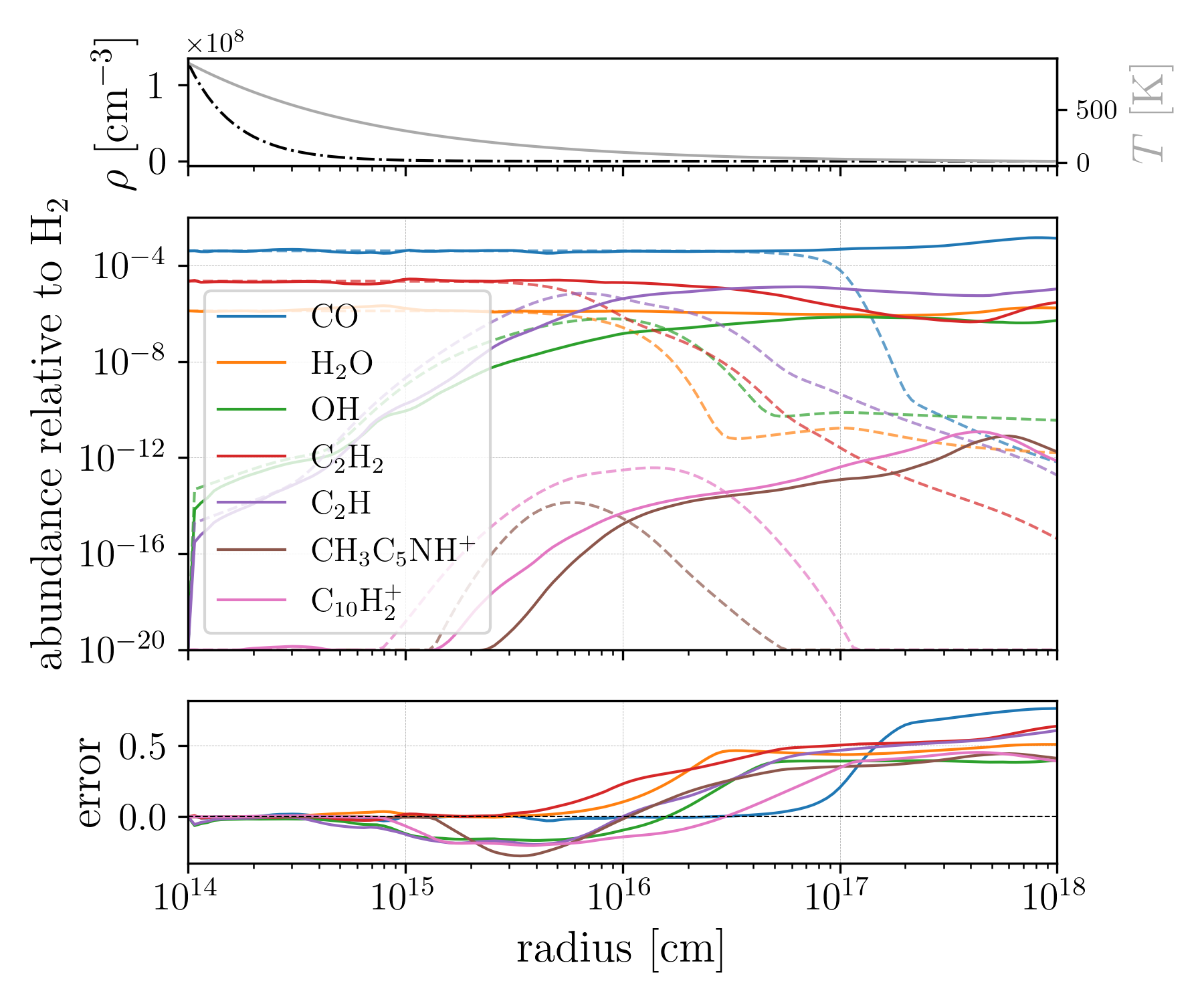}
    \includegraphics[width=0.475\textwidth]{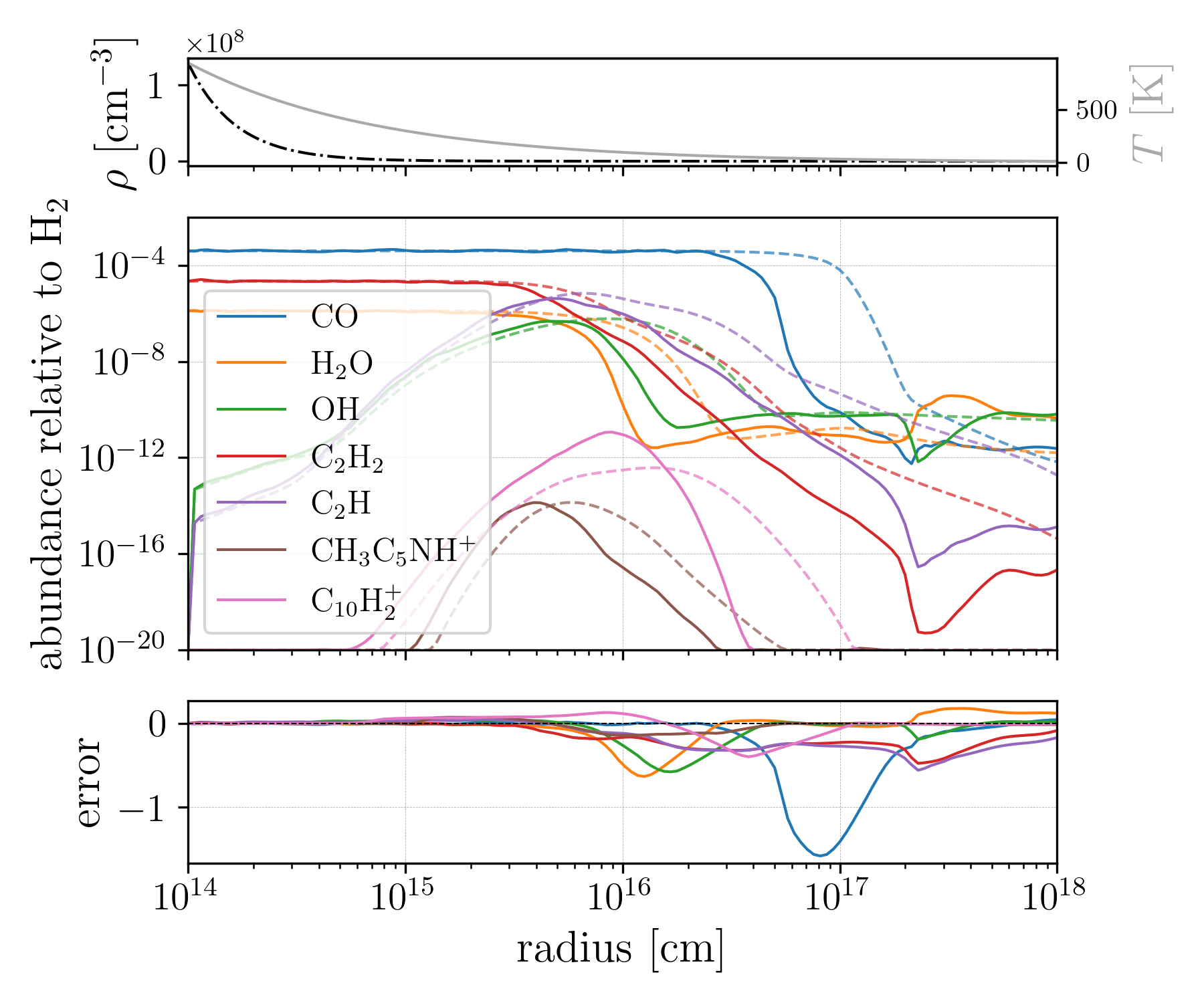}
    \caption{Evolution test of model $\ell oc$\emph{1} (\underline{left}) and $\ell oc$\emph{4} (\underline{right}) on a \csone\ models with the following input parameters: $\Mdot=1\times10^{-6}\,\Msolyr$, $\vexp=17.5\,\kms$, $\Tstar=2300\,$K, and $\eps=0.55$. \underline{Upper panel:} H$_2$ number density (dashed-dotted, left $y$-axis) and temperature (full grey, right $y$-axis) as a function of outflow radius. \underline{Middle panel:} Abundances of specific species, given in legend. The dashed line gives the result of the classical model (ground truth), the full line gives the result of \mace. \underline{Lower panel:} Error (Eq.\ \ref{eq:errorlog}) of the \mace\ model compared to the classical model.}
    \label{fig:ell1_4}
\end{figure}

In this section, we show more predicted abundance profiles by the \mace\ models, trained according to the local scheme (Eq.\ \ref{eq:totloss_loc} and Table \ref{tab:local-models}). Fig.\ \ref{fig:ell1_4} shows model $\ell oc$\emph{1} (left) and $\ell oc$\emph{4} (right) apply on a \csone\ test model.

\section{Testing integrated models}\label{sect:integrated_app}
\begin{figure}[h]
    \centering
    \includegraphics[width=0.475\textwidth]{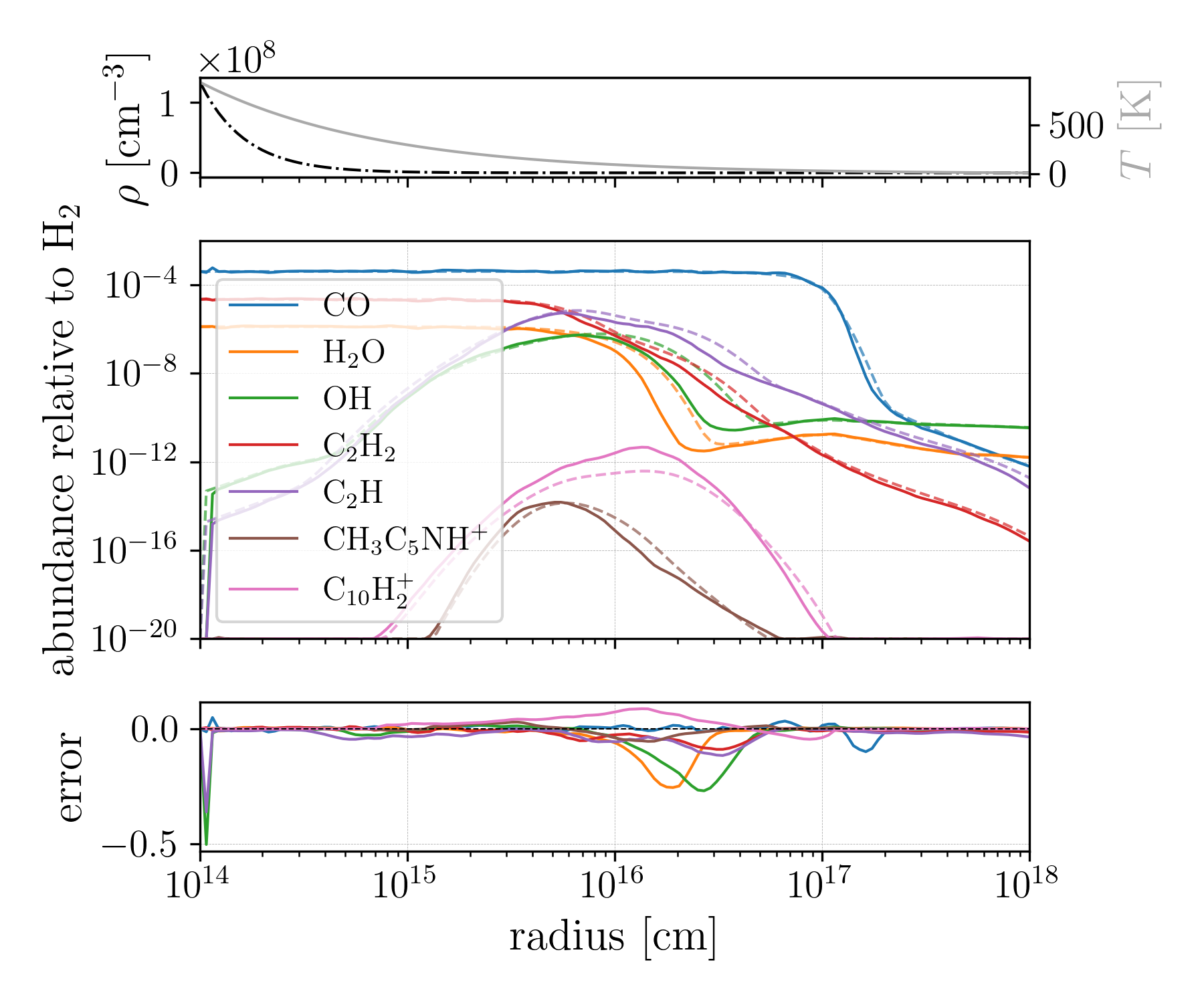}
    \includegraphics[width=0.475\textwidth]{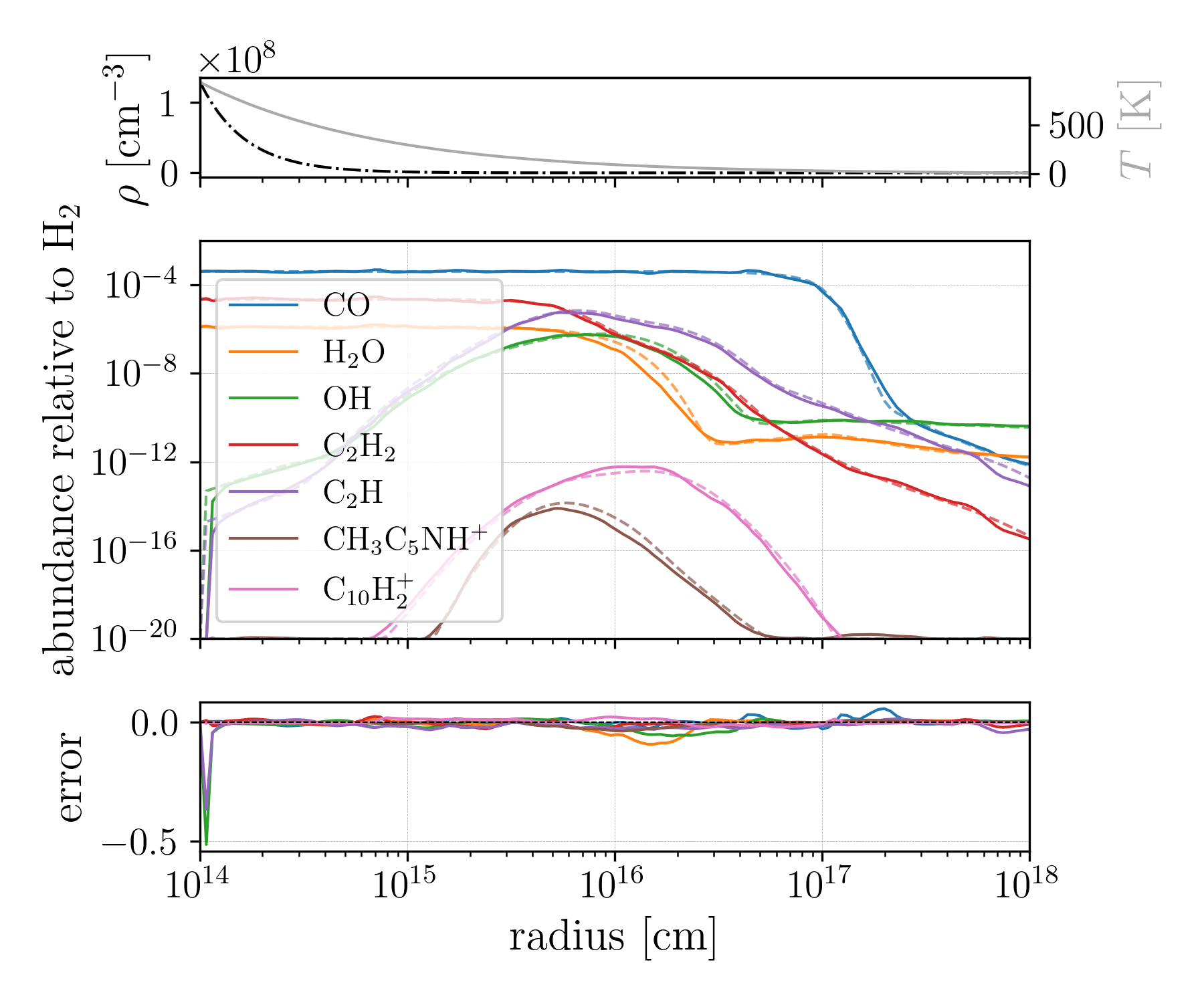}
    \caption{Evolution test of model \emph{int2} (\underline{left}) and \emph{int7} (\underline{right}) on a \csone\ models with the following input parameters: $\Mdot=1\times10^{-6}\,\Msolyr$, $\vexp=17.5\,\kms$, $\Tstar=2300\,$K, and $\eps=0.55$. \underline{Upper panel:} H$_2$ number density (dashed-dotted, left $y$-axis) and temperature (full grey, right $y$-axis) as a function of outflow radius. \underline{Middle panel:} Abundances of specific species, given in legend. The dashed line gives the result of the classical model (ground truth), the full line gives the result of \mace. \underline{Lower panel:} Error (Eq.\ \ref{eq:errorlog}) of the \mace\ model compared to the classical model.}
    \label{fig:int1_4}
\end{figure}

\begin{figure}[h]
    \centering
    \includegraphics[width=0.475\textwidth]{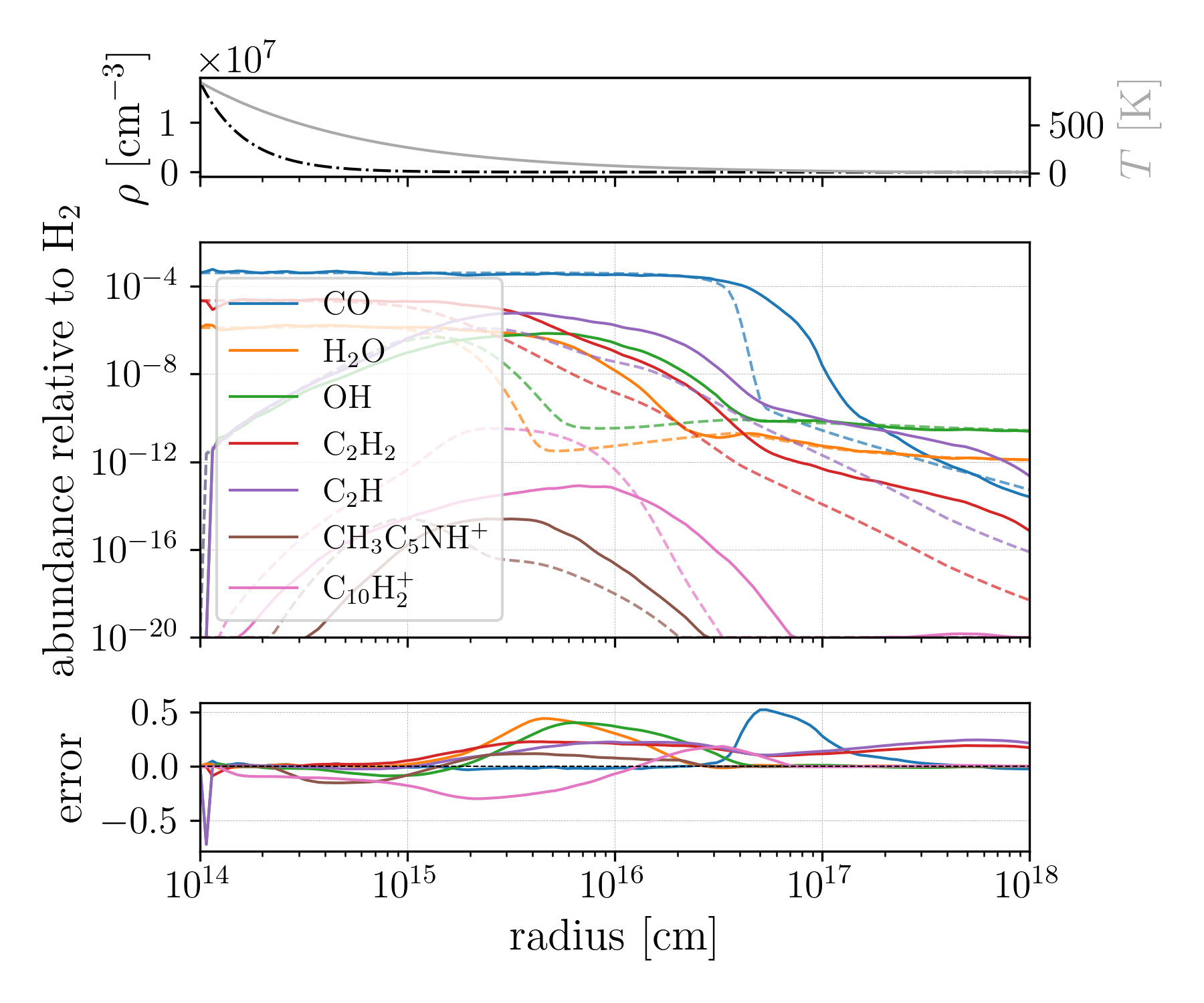}
    \includegraphics[width=0.475\textwidth]{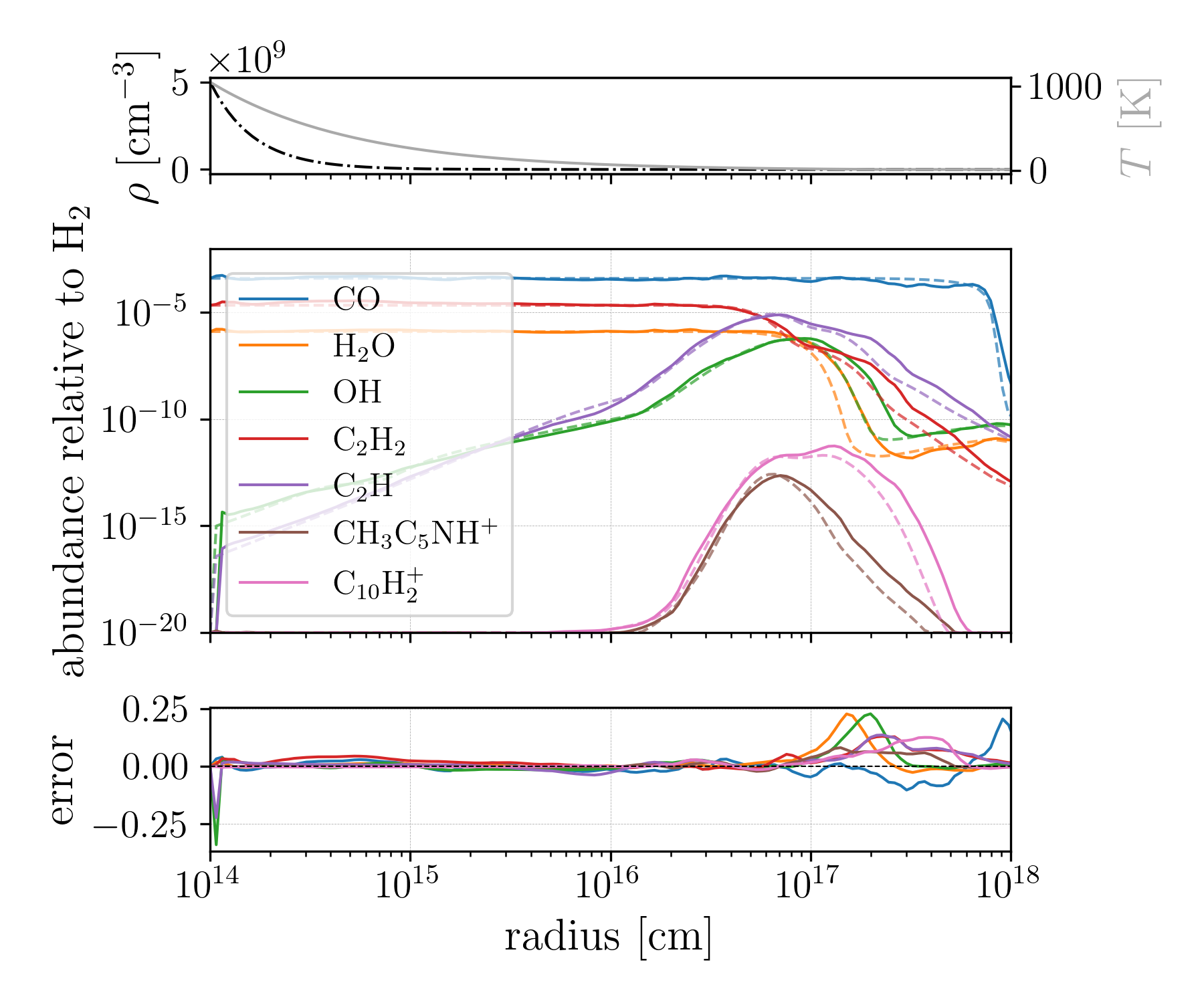}
    \caption{Evolution tests of model \emph{int4}. \underline{Left:} low-density \csone\ models with input parameters $\Mdot=2\times10^{-8}\,\Msolyr$, $\vexp=2.5\,\kms$, $\Tstar=2100\,$K, and $\eps=0.5$. \underline{Right:} high-density \csone\ models with input parameters $\Mdot=5\times10^{-5}\,\Msolyr$, $\vexp=22.5\,\kms$, $\Tstar=2750\,$K, and $\eps=0.6$. \underline{Upper panel:} H$_2$ number density (dashed-dotted, left $y$-axis) and temperature (full grey, right $y$-axis) as a function of outflow radius. \underline{Middle panels:} Abundances of specific species, given in legend. The dashed line gives the result of the classical model (ground truth), the full line gives the result of \mace. \underline{Lower panels:} Error (Eq.\ \ref{eq:errorlog}) of the \mace\ model compared to the classical model.}
    \label{fig:int4_low}
\end{figure}

In this section, we show more predicted abundance profiles by the \mace\ models, trained using the integrated scheme (Eq.\ \ref{eq:totloss_int} and Table \ref{tab:int-models}). Fig.\ \ref{fig:int1_4} shows the predicted abundance profile by models \emph{int2} (left) and \emph{int7} (right). Fig.\ \ref{fig:int4_low} shows tests of model \emph{int4} on a low- (left) and high-density (right) outflow.

% \clearpage

\FloatBarrier
\section{Implementing element conservation}\label{sect:elementcons}
In this section, we elaborate on the implementation of element conservation in the emulator. The matrix $M_{Ii}$ defines how much of each element, $I$, appears in each chemical species, $i$, of the considered chemical network. The elemental abundance $e_I$ then yields $M_{Ii} \, n_{i}$ and should be conserved, hence
\begin{equation}
    M_{Ii} \frac{{\text{d}} n_{i}}{\text{d}t} =  0.
\end{equation}
Given (i) that the decoder maps latent abundances to real abundances, $n_{i} \ = \ \mathcal{D}_{i}(\boldsymbol{z})$ (Eq.\ \ref{eq:decoder}) and (ii) the dynamics in latent space (Eq.\ \ref{eq:latentODE}), we can rewrite the dynamics in real space as
\begin{equation}
    \frac{\text{d} n_{i}}{\text{d} t}
     = 
    \partial_{\alpha} n_{i}    
    \frac{\text{d} z_{\alpha}}{\text{d} t}
     = 
    \partial_{\alpha} \mathcal{D}_{i} (\boldsymbol{z})
    \frac{\text{d} z_{\alpha}}{\text{d} t}
     = 
    \partial_{\alpha} \mathcal{D}_{i} (\boldsymbol{z}) 
    \big(
        \mathcal{C}_{\alpha}  + \ \mathcal{A}_{\alpha \beta}  z_{\beta}  +  \mathcal{B}_{\alpha \beta \gamma}  z_{\beta}  z_{\gamma}
    \big) .
\end{equation}
Using the dynamics in real space, conservation of elemental abundance (Eq.\ \ref{eq:latentODE}) thus implies
\begin{equation}\label{eq:elementcons}
    M_{Ii} 
    \partial_{\alpha} \mathcal{D}_{i} (\boldsymbol{z}) 
    \big(
        \mathcal{C}_{\alpha} +  \mathcal{A}_{\alpha \beta}  z_{\beta}  +  \mathcal{B}_{\alpha \beta \gamma}  z_{\beta}  z_{\gamma}
    \big)  =  0,
\end{equation}
with $\partial_{\alpha} = \partial / \partial z_{\alpha}$.
\\ \indent
The conservation of elemental abundance can be implemented in the emulator in two ways. (i) The conservation of elements can be present as an additional loss term to the total loss. Hence, the element loss (ELM) would be defined as
\begin{equation}
    L_{\rm ELM} = M_{Ii} 
    \partial_{\alpha} \mathcal{D}_{i} (\boldsymbol{z}) 
    \big(
        \mathcal{C}_{\alpha}  +  \mathcal{A}_{\alpha \beta}  z_{\beta}  +  \mathcal{B}_{\alpha \beta \gamma}  z_{\beta}  z_{\gamma}
    \big).
\end{equation}
We have tried implementing this element loss in \mace. However, this slowed down the training immensely, because for every pass of training data through \mace, the jacobian of the decoder neural network $\partial\mathcal{D}(\z)$ needs to be calculated and multiplied by the tensor coefficients of the latent ODE, which involves many operations on large matrices. (ii) The conservation of elements can be incorporated explicitly in the architecture itself, by constructing the latent ODE (Eq.\ \ref{eq:latentODE}) in such a way that Eq.\ \eqref{eq:elementcons} is always satisfied. This can be done in many different ways, requiring many explicit design choices, such as which parameters to determine using this constraint, which is beyond the scope of this paper.

% Appendices can be broken into separate sections just like in the main text.
% The only difference is that each appendix section is indexed by a letter
% (A, B, C, etc.) instead of a number.  Likewise numbered equations have
% the section letter appended.  Here is an equation as an example.
% \begin{equation}
% I = \frac{1}{1 + d_{1}^{P (1 + d_{2} )}}
% \end{equation}
% Appendix tables and figures should not be numbered like equations. Instead
% they should continue the sequence from the main article body.


\bibliography{biblio.bib}{}
\bibliographystyle{aasjournal}



\end{document}